\documentclass[aps,prl,amsmath,showpacs,preprintnumbers,twocolumn,amssymb,superscriptaddress]{revtex4-1}

\usepackage{graphicx}
\usepackage{mathptmx, textcomp}
\usepackage[latin1]{inputenc}
\usepackage{tabularx}
\usepackage{units}
\usepackage{color}
\definecolor{bl}{rgb}{0, .1, .6}
\definecolor{rd}{rgb}{1,0,.2}

\usepackage[colorlinks=true, citecolor = bl, linkcolor = bl, urlcolor=bl, pdfborder={0 0 0}]{hyperref}
\usepackage{bibunits}

\newcommand{\muK}{\mu {\rm K}}
\newcommand{\kb}{k_{\rm B}}

\newcommand{\eb}{E_{\rm b}}
\newcommand{\MHz}{{\rm MHz}}
\newcommand{\kHz}{{\rm kHz}}
\newcommand{\Hz}{{\rm Hz}}
\newcommand{\g}{{\rm G}}
\newcommand{\mg}{{\rm mG}}
\newcommand{\ms}{{\rm ms}}
\newcommand{\be}{\begin{eqnarray}}
\newcommand{\ee}{\end{eqnarray}}
\newcommand{\abg}{a_{\rm bg}}
\newcommand{\abgt}{\tilde{a}_{\rm bg}}
\newcommand{\sres}{s_{\rm res}}
\newcommand{\ldb}{\lambda_{\rm dB}}

\newcommand{\den}{{\rm cm}^{-3}}

\begin{document}
\begin{bibunit}[apsrev4-1]
\title{Broad universal Feshbach resonances in the chaotic spectrum of Dysprosium atoms}

\author{T. Maier}
\altaffiliation{These authors contributed equally to this work.}
\affiliation{5. Physikalisches Institut and Center for Integrated Quantum Science and Technology,
Universit\"at Stuttgart, Pfaffenwaldring 57, 70550 Stuttgart, Germany}
\author{I. Ferrier-Barbut}
\altaffiliation{These authors contributed equally to this work.}
\affiliation{5. Physikalisches Institut and Center for Integrated Quantum Science and Technology,
Universit\"at Stuttgart, Pfaffenwaldring 57, 70550 Stuttgart, Germany}
\author{H. Kadau}
\affiliation{5. Physikalisches Institut and Center for Integrated Quantum Science and Technology,
Universit\"at Stuttgart, Pfaffenwaldring 57, 70550 Stuttgart, Germany}
\author{M. Schmitt}
\affiliation{5. Physikalisches Institut and Center for Integrated Quantum Science and Technology,
Universit\"at Stuttgart, Pfaffenwaldring 57, 70550 Stuttgart, Germany}
\author{M. Wenzel}
\affiliation{5. Physikalisches Institut and Center for Integrated Quantum Science and Technology,
Universit\"at Stuttgart, Pfaffenwaldring 57, 70550 Stuttgart, Germany}
\author{C. Wink}
\affiliation{5. Physikalisches Institut and Center for Integrated Quantum Science and Technology,
Universit\"at Stuttgart, Pfaffenwaldring 57, 70550 Stuttgart, Germany}
\author{T. Pfau}
\affiliation{5. Physikalisches Institut and Center for Integrated Quantum Science and Technology,
Universit\"at Stuttgart, Pfaffenwaldring 57, 70550 Stuttgart, Germany}
\author{K. Jachymski}
\affiliation{Faculty of Physics, University of Warsaw, Pasteura 5, 02-093 Warsaw, Poland}
\author{P. S. Julienne}
\affiliation{Joint Quantum Institute, University of Maryland and National Institute of Standards and Technology, College Park, Maryland 20742, USA}

\date{\today}


\begin{abstract}
We report on the observation of weakly-bound dimers of bosonic Dysprosium with a strong universal s-wave halo character, associated with broad magnetic Feshbach resonances. These states surprisingly decouple from the chaotic backgound of narrow resonances, persisting across many such narrow resonances. In addition they show the highest reported magnetic moment $\mu\simeq20\,\mu_{\rm B}$ of any ultracold molecule. We analyze our findings using a coupled-channel theory taking into account the short range van der Waals interaction and a correction due to the strong dipole moment of Dysprosium. We are able to extract the scattering length as a function of magnetic field associated with these resonances and obtain a background scattering length $\abg=91(16)\,a_0$. These results offer prospects of a tunability of the interactions in Dysprosium, which we illustrate by observing the saturation of three-body losses. 

\end{abstract} 

\maketitle

Magnetic Feshbach resonances have emerged as an essential tool for engineering interactions in atomic and molecular physics \cite{Chin:2008}. On the one hand they have been used to study genuine many-body systems in quantum gases, such as ultracold fermions in the BEC to BCS crossover \cite{Zwerger:2012}. On the other hand they have allowed major advances in few-body physics, through the study of the Efimov effect \cite{Ferlaino:2011}, or the production of ground state molecules \cite{Danzl:2008,Ni:2008}. The continuous progress in cooling and control of atoms and molecules now opens the door to the study of novel condensed-matter systems using species with highly complex spectra. However the complexity of the tunable Feshbach resonance spectrum of such species naturally raises the question as to whether precise control of interactions will be possible while avoiding harmful inelastic loss collisions. An illustration of this is found in the heavy, sub-merged shell elements Dysprosium and Erbium, both of which have been brought to quantum degeneracy \cite{Aikawa:2012, Aikawa:2014, Lu:2011, Lu:2012}. Their complex electronic structure leads both to a large magnetic dipole moment and to a strong anisotropy of the van der Waals interaction, which give rise in turn to an extremely dense spectrum of very narrow Feshbach resonances, characterized by a chaotic distribution \cite{Frisch:2014,Maier:2015}. In order to use such lanthanide species to study long-range many-body interacting systems, it is essential to understand and be able to control their collisional properties.

In this Letter we demonstrate that broad Feshbach resonances actually exist for the lowest-lying state of $^{164}$Dy, $J=8, m_J=-8$. Furhermore these resonances decouple from the chaotic background and thus offer a broad tunability of interactions that should be readily useable for few- and many-body studies. We present Feshbach spectroscopy measurements up to a magnetic field of $600\,\g$. This spectroscopy reveals loss features with widths of the order of several Gauss. Magnetic field modulation spectroscopy allows us to measure the binding energy of weakly bound dimers. In two cases we identify a bound state that has a persistent universal s-wave halo character over several Gauss despite crossing with a large number of other bound states.

\begin{figure*}[htbp]
\includegraphics[width=\textwidth]{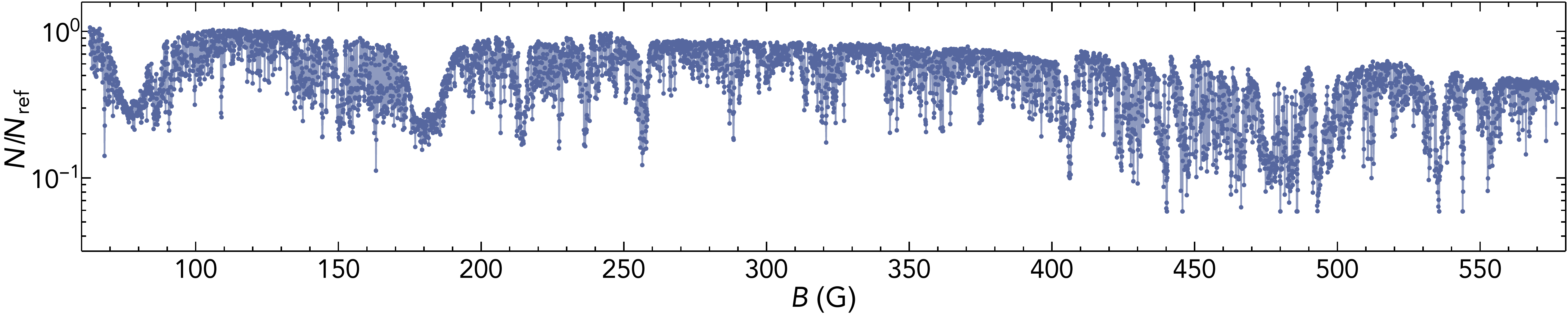}
\caption{Atom-loss spectroscopy at a temperature of $2.4\,\muK$ mapping the Feshbach spectrum of $^{164}\text{Dy}$ in $J=8,\,m_J=-8$ with 100$\,$mG resolution. The atom number is normalized to a reference number $N_{\rm ref}$ taken at low field every 30 images.}
\label{fig:resonances}
\end{figure*}

\begin{figure*}[htbp]
\includegraphics[width=\textwidth]{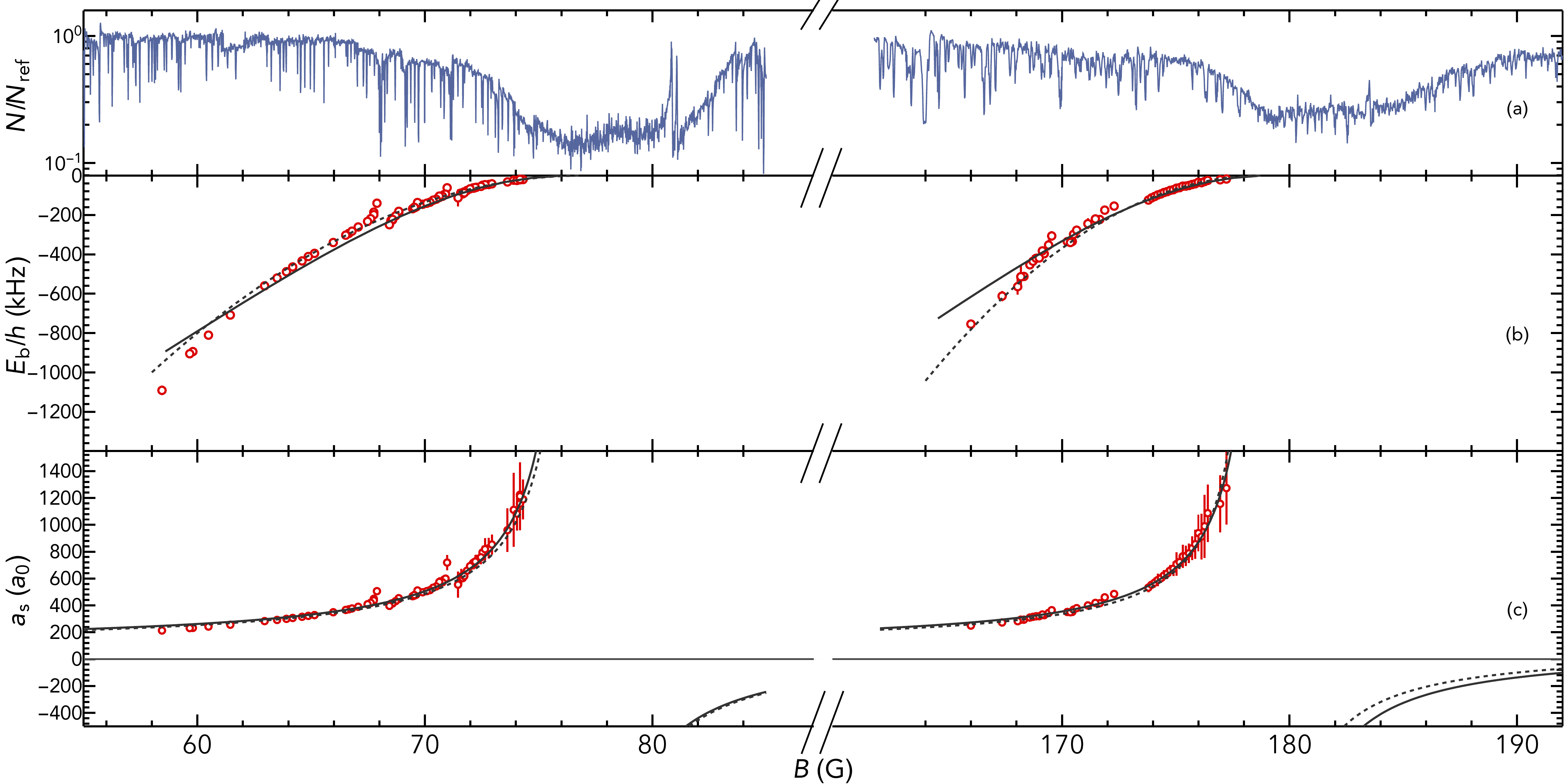}
\caption{(color online) Broad Feshbach resonances of $^{164}$Dy. (a) Atom-loss spectroscopy at $T=500\,{\rm nK}$ with a resolution of $20\,\mg$. This spectrum shows one broad feature overlapping with many narrow resonances. (b) Binding energy of weakly bound dimers measured by magnetic field modulation spectroscopy (red circles). The solid (dashed) lines are obtained by fits of our data to the coupled-channel calculations (universal expression), from these fits we extract $\abg\,\Delta$ (table 1). (c) The red circles are obtained by converting the $\eb(B)$ data using the coupled-channel calculations for $a(\eb)$. The sold lines are a fit to this data using eq.~\ref{eq:aofb}. The dashed lines are the scattering length resulting from the fit of $\eb(B)$ by the universal expression assuming $\abg=91\,a_0$ (see main text). }
\label{fig:bindingenergy}
\end{figure*}

To study Feshbach resonances we prepare ultracold samples of $^{164}\text{Dy}$ atoms in their lowest Zeeman sublevel $m_J=-8$, with a magnetic moment $\mu_{\rm at}=9.93\,\mu_{\rm B}$ where $\mu_{\rm B}$ is the Bohr magneton, the details of our preparation methods are presented in \cite{Maier:2014,Maier:2015}. The magnetic field was calibrated via radio-frequency (RF) spectroscopy between the two lowest Zeeman sub-levels of $^{164}$Dy. Feshbach spectroscopy is performed by ramping up a magnetic offset field in 15$\,$ms to the desired target value in the range from $60\,\g$ to $600\,\g$. Typical samples contain $10^5$ atoms at a temperature of $2.4\,\muK$ (see supplemental material \cite{SupMat} for details). The atoms are held for a wait time ($2\,{\rm s}$ in fig.~\ref{fig:resonances}) in a constant optical dipole trap where they undergo inelastic collisions inducing losses. 
Figure~\ref{fig:resonances} shows the final atom number measured after a time-of-flight as a function of magnetic field with a resolution of $100\,\mg$, mapping the Feshbach resonance spectrum. In this spectrum, an irregular pattern of several broad features appears on top of the sea of narrow resonances.
Among them we identify in particular two distinct broad loss features located near 80$\,$G and 180$\,$G. In the following, we focus on these two features.
A magnetic field scan with fine resolution ($20\,\mg$) over these two resonances shows that inside the broad features, the dense background of narrow resonances remain (see figure~\ref{fig:bindingenergy}, and \cite{SupMat}). 

It is usually accepted that broad features in atom-loss spectra correspond to single broad resonances \cite{Inouye:1998,Vuletic:1999,Repp:2013,Tung:2013}. However, in the chaotic Feshbach spectrum of lanthanides, this correspondence is not straightforward, and the large number of interaction potentials prevents so far the association of a loss resonance to a particular bound state, except in a few cases at very low magnetic field as shown by Ref.~\cite{Frisch:2015}. To circumvent this shortcoming, we measure the bound states energy versus magnetic field using the standard technique of molecular association by magnetic field modulation \cite{Thompson:2005,Langmack:2014}. We implement this spectroscopy on thermal clouds at a temperature $T$ with $400\,{\rm nK}\leqslant T\leqslant700\,{\rm nK}$, and focus on the regions $B\in[55\,\g,75\,\g]$, and $B\in[160\,\g,180\,\g]$. The magnetic field is modulated around its bias value $B$ at radio frequencies during a time between $100\,{\rm ms}$ and $500\,{\rm ms}$ and with a modulation amplitude lying between $100\, {\rm mG}$ and $500\, {\rm mG}$.

With this spectroscopy we are able to observe the states responsible for narrow resonances but we also observe an isolated loss resonance with typical FWHM $<30\,\kHz$. This feature appears only on the low-field side of the two broad features at $B\simeq80\,\g$ and $B\simeq180\,\g$ and we associate it with weakly bound dimers with slowly-varying binding energy $\eb$. We extract $\eb(B)$ \cite{SupMat} and plot it in red dots in figure \ref{fig:bindingenergy} (b). We put this in perspective with high-resolution atom-loss scans (a). In both cases we observe a very slow variation of $\eb$ with magnetic field ($<1\,\MHz$ over $10\,\g$). This implies that these molecules have a magnetic moment $\mu$ very close to that of two free atoms, $\mu=2\mu_{\rm at}-\frac{d\eb}{dB}\simeq20\,\mu_{\rm B}$. This value is the highest reported for a magnetic dipole \cite{Frisch:2015}. Furthermore, we observe a a quadratic behaviour for $\eb(B)$ over several gauss, which reveals a coupling of the bound state with the two-atom continuum, conventionally termed the open channel \cite{Chin:2008}. This constitutes a strong evidence that the observed bound state is in both cases the s-wave halo state that is found in the vicinity of Feshbach resonances. It is striking that this strong open channel character is found over a broad field range, despite the crossing of this state with many other bound states which are responsible for the narrow resonances dense background. In fact, near the lower resonance we observe signatures of avoided crossings with other bound states, visible in Fig. \ref{fig:bindingenergy}, which correlate with narrow resonances in the atom-loss spectrum.

We now provide a theoretical basis supporting the existence of such halo states. We base the theoretical analysis (detailed in~\cite{SupMat}) of these observations on the dipole-modified $s$-wave scattering model in which a strong, open-channel-dominated resonance overlaps with many weak, narrow resonances. Our observations provide clear evidence that the effect of such a broad resonance persists across many narrow features and that away from narrow poles the scattering length $a$ takes the usual expression
\be
a(B)=\abg\left(1-\frac{\Delta}{B-B_0}\right).\label{eq:aofb}
\ee 
where $B_0$ is the resonance pole, $\Delta$ its width and $\abg$ the background scattering length.
Furthermore in the pole vicinity, binding energies $\eb \lesssim 1$ MHz are well approximated by the universal expression: 
\be
\eb=-\frac{\hbar^2}{ma^2}\label{eq:ebofa}
\ee
with $m$ is the mass of a Dysprosium atom and $\hbar$ Planck's constant. 

In our theoretical model we account for slight deviations from the universal relation~\eqref{eq:ebofa} that result from the form of the long-range interaction potential which contains a typical van der Waals $-C_6/r^6$ term and a contribution from dipole-dipole interactions which for the $s$-wave has $\propto 1/r^4$ character~\cite{Bohn:2009}. We construct a two-channel model with short-range coupling to a ramping bound state in the closed channel. The model parameters are $\abg$, $B_0$, the magnetic moment difference between channel states $\delta\mu$, and the dimensionless pole strength $\sres\propto \abg\Delta\delta\mu$ \cite{Chin:2008,SupMat}. In general, large $\sres$ values indicate open-channel-dominated resonances with highly universal behavior. 

We use this coupled-channel (CC) model to describe the observed data. We found out that, as expected, the observed resonances have a high $\sres$. The data allows us to extract only a lower bound $\sres\gtrsim10$ and we cannot obtain $\delta\mu$ either. Rather we can only extract the product $\abg\,\Delta$. We obtain $\abg\,\Delta$ by a direct fit to our $\eb(B)$ data \cite{SupMat}, the result of which is represented in Fig.\ref{fig:bindingenergy} (b) as a solid line. Second, we calculate the binding energy dependence on the scattering length $\eb(a)$ assuming that $\abg$ is close to $100\,a_0$, which will be justified below. We use it to translate the experimental binding energies to scattering lengths (red circles in Fig.~\ref{fig:bindingenergy} (c)). Comparing the result to~\eqref{eq:aofb} also allows then to extract $\abg\,\Delta$. As expected these two methods agree, with a fit result given in Table~\ref{tabres}. In addition we also simply fit our $\eb(B)$ data by the universal quadratic dependence ($\eb=-(B-B_0)^2/(\abg\Delta)^2$) in the vicinity of the pole ($B>66\,\g$ and $B>173\,\g$ for the two resonances respectively), with a fit result shown in dashed line in Fig.~\ref{fig:bindingenergy} (b) and in Table~\ref{tabres}. The coupled-channel theory and the universal fit are in close agreement with each other for the lower field resonance and in reasonable agreement for the higher-field one. We assume $\abg=91\,a_0$ (as justified below) to calculate $a(B)$ from the value of $\abg\Delta$ given by the universal fit, in close agreement with the CC results. Our measure of $a$ is valid when the magnetic field is tuned away from the narrow resonances as is visible in Fig.~\ref{fig:bindingenergy} (c) (see also~\cite{SupMat} for a close zoom).

The $\eb(B)$ data displays a quadratic field dependence well beyond the pole vicinity. This indicates that the observed broad resonances have decisive impact on the scattering properties over a large magnetic field region, although multiple couplings to other states can be found in the data. The investigation of the coupling between the halo-state and other bound states will be the topic of future research.

\begin{table}
\begin{tabular}{|c|c|c|c|}
\hline 
 universal $B_0$ & universal $\abg \Delta$ & numerical CC $B_0$& numerical CC $\abg \Delta$\\
\hline \hline
 $76.9(5)\,\g$ & 2810(100)$\,\g\, a_0$ & $76.8(5)\,\g$ & 2700(100)$\,\g\, a_0$\\
 $178.8(6)\,\g$ & 2150(150)$\,\g\, a_0$ & $179.1(6)\,\g$ & 2540(110)$\,\g\, a_0$\\ 
\hline
\end{tabular}
\caption{\label{tabres}Resonance parameters obtained from fitting the binding energy data to the universal formula~\eqref{eq:ebofa} and to numerical coupled channels calculation.}
\end{table}

Furthermore we observe several indications that the broad resonance provides a background scattering length for the narrow resonances (see \cite{SupMat}) that indeed crosses  $0$ on the high-field side of the $76.9\,\g$ resonance. For non-magnetic atoms, a vanishing $a$ results in an absence of elastic collisions at low temperature, rendering evaporation ineffective \cite{Jochim:2002}. As a consequence, the final atom number and temperature after a given wait time are maximal. In the present case of Dy, though $a$ vanishes, the dipolar interaction still induces collisions characterized by the dipolar length $D=m\mu_0\mu_{\rm at}^2/8\pi\hbar^2=196\,a_0$ and the collisional cross section $\sigma_{\rm dip}=32\pi D^2/45$ \cite{Bohn:2009}, \footnote{We have $D = 3/2 a_{\rm dd}$ where $a_{\rm dd}$ is defined in \cite{Lahaye:2009}.}. Nonetheless when $a=0$, the cross section and then thermalization is minimal. This leads to a slower evaporation, characterized by a maximum in the background (in-between narrow resonances) atom number and temperature after a holding time in a constant trap. We do observe a maximum in these two observables, located for the atom number at $B_{{\rm Max,}N}=109\,(5)\,\g$ and for temperature at $B_{{\rm Max,}T}=107\,(5)\,\g$ (see \cite{SupMat} for a detailed figure). We join this observation with a second analysis, where we make use of the asymmetric lineshape of the narrow resonances. Indeed for a narrow resonance, the sign of its ``local'' background scattering length $\abgt$~\cite{Jachymski:2013,SupMat} influences the symmetry of its lineshape in the atom-loss spectrum. If $\abgt>0$ ($\abgt<0$) the local zero crossing attached to this resonance is on the high-(low-)field side of its pole, and the three-body loss spectrum is expected to show a minimum near such zero crossing. Most of the resonances we observe are too narrow to asses the symmetry of their lineshape given our resolution, however we can still observe several ones at different fields, and they show show that below the $76.9\,\g$ resonance, $\abgt>0$. For higher fields we observe $\abgt<0$, up to a field of about $109\,\g$ where we actually observe a broad enough resonance, with a very symmetric lineshape, corresponding to $\abgt$ close to zero (see \cite{SupMat} for spectra). This joint analysis shows that there is an overall ``background'' scattering length that reaches 0 at a field $B=108\,(5)\g$. Assuming that the broad $s$-wave halo resonance continues to provide the local B-dependent background to the narrow resonances~\cite{Jachymski:2013}, the universal model~\eqref{eq:aofb} then implies a width $\Delta=$31(6) G and $\abg=$91(15) $a_0$ for this resonance, justifying the assumptions made above. This value is also in agreement with the rethermalization measurements of ref.~\cite{Tang:2015} performed at low fields. While the data suggests the persistence of such strong $s$-wave halo influence across a set of narrow resonances consistently with~\cite{Jachymski:2013}, continuing experimental and theoretical work is needed to explore this effect.

\begin{figure}[t]
\includegraphics[width=\columnwidth]{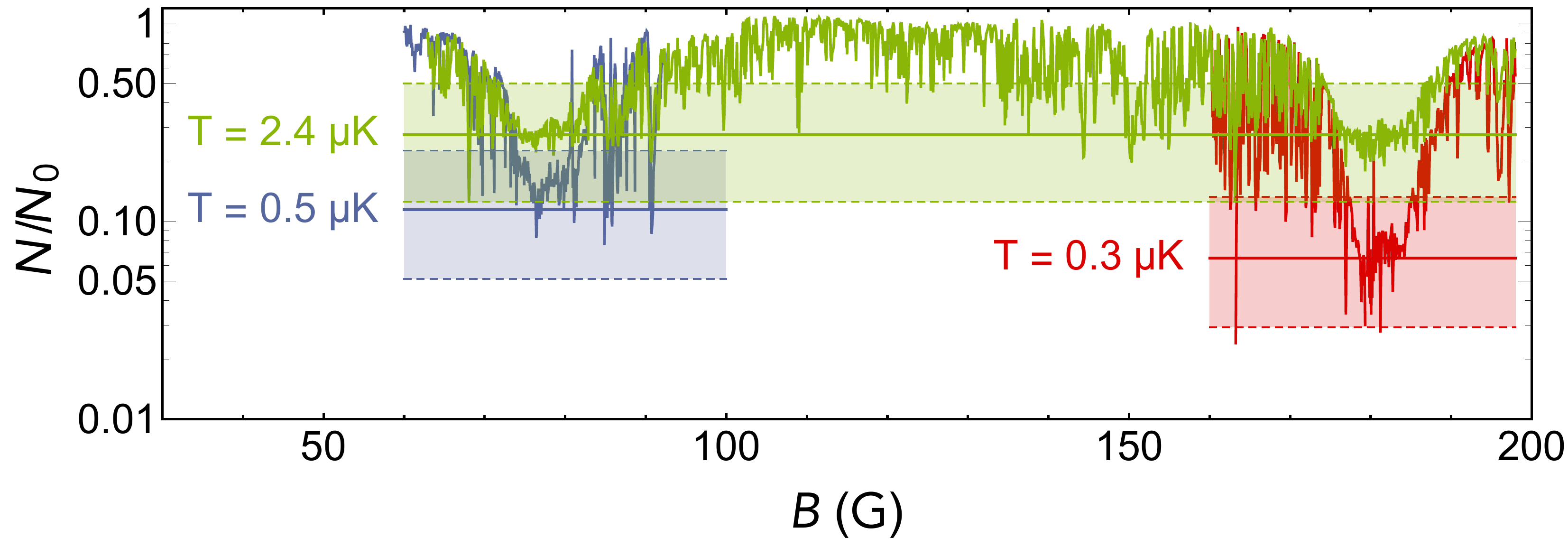}
\caption{(color online) Atom loss spectroscopy for different experimental starting conditions and wait times. Final atom number $N$ normalized to the initial one $N_0$ as a function of magnetic field, with a resolution of $100\,\mg$. Green: initial temperature: $T_0=2.4\,\muK$, initial density: $n_0=6.1(2.0)\times10^{12}\,\den$ wait time $t=2s$, blue: $T_0=0.5\,\muK$, $n_0=3.7(1.3)\times10^{12}\,\den$, $t=500\,\ms$, red: $T_0=0.3\,\muK$, $n_0=4.6(1.6)\times10^{12}\,\den$  $t=300\,\ms$. 
The temperature dependence of the saturation is well reproduced by the model of universal loss dynamics of unitary Bose gases of ref.~\cite{Eismann:2015} (solid horizontal lines). The shaded regions represent the uncertainty on the results of the model given a one-standard deviation on all experimental parameters entering the model (temperature, atom number, trap frequency and depth).}
\label{fig:losses}
\end{figure}

We have characterized two Feshbach resonances with $\sres\gg1$, we thus expect to observe effects characteristic for broad resonances. We now show evidence that our observations of atom losses in the vicinity of the poles are indeed compatible with universal loss dynamics found at the center of broad resonances. For this we study atom-losses at different temperatures. We observe that approaching the poles, the final atom number reaches a minimum that is the same on the two broad resonances. Furthermore, at lower temperature, the saturation is reached in a narrower field region and at a lower level despite a smaller central density in the initial conditions, Fig.~\ref{fig:losses}. 

The temperature dependence of three-body losses of Bose gases is known in the unitary regime where $a/\ldb\gg1$, with $\ldb$ the thermal de Broglie wavelength. For a unitary Bose gas, the three-body loss parameter $L_3$ takes a $1/T^2$ behaviour: 
$L_3= \frac{\hbar^5}{m^{3}}\,36\sqrt{3}\,\pi^2 \frac{1-e^{-4\eta_*}}{(k_B T)^{2}}$ \cite{Rem:2013}, \footnote{A 7\% modulation of $L_3$ for identical bosons is omitted in this expression, see \cite{Rem:2013}.}. The elasticity parameter $\eta_*$ characterizes the efficiency with which three atoms in contact recombine to a dimer and a free atom.

We compare our findings with the loss dynamics model developed in \cite{Eismann:2015}, which predicts the final atom number taking into account two-body evaporation and three-body recombination. This model requires the knowledge of the trap depth, frequencies and initial atom number and temperature. These observables are calibrated (see \cite{SupMat}), and the only unknown parameter of this model is $\eta_*$. We then compare this model with a fixed $\eta_*$ to the three data sets represented in figure~\ref{fig:losses}. Within experimental uncertainties, we find that this model reproduces well our final atom number at resonance for a fixed $\eta_*$ that we find to be
\be
\eta_*=0.07^{+0.17}_{-0.05}.
\ee
This value is close to the lowest reported values for alkali-metal atoms \cite{Kraemer:2006,Zaccanti:2009,Fletcher:2013,Gross:2010,Dyke:2013,Wild:2012} and implies lifetimes comparable to these species in the unitary regime. Having this experimental information is valuable since no theoretical prediction is available. We observe that $\eta_*$ is identical for our two broad resonances. A systematic study of $L_3(T)$ at resonance would yield a more precise measure.\\

In conclusion, it is remarkable that despite an apparent chaotic distribution of Feshbach resonances, isolated states can decouple from this background and lead to a simple description. This emergence of structure from a chaotic background is a generic feature of quantum-chaotic systems \cite{Heller:1984}, with the emergent states associated with classical quasiperiodic trajectories in phase space \cite{Jacobson:1999}. Further work is needed to understand the origin of our broad resonances. However the simple universal description provided here is sufficient to predict how to control dipolar gases in an understood way \cite{Lahaye:2007}. Between narrow resonances we obtain pure BECs with up to $25\times 10^3$ atoms with long lifetimes ($\gtrsim1\,{\rm s}$). The scattering length obtained in this work can thus be confirmed with precision by the measurement of the density distribution of Bose-Einstein condensates. Such measurements are ongoing with our setup. This opens prospects of studying the few-body physics of bosonic dipoles; for instance the energy of Efimov states is thought to be impacted by the dipole-dipole interaction \cite{Wang:2011}.\\

\begin{acknowledgments}

We thank Axel Griesmaier for support at the early stage of the experiment. This work is supported by the German Research Foundation (DFG) within SFB/TRR21. H.K. acknowledges support by the 'Studienstiftung des deutschen Volkes'. K.J. acknowledges financial support by the Foundation for Polish Science International PhD Projects Programme co-financed by the EU European Regional Development Fund. PSJ acknowledges support by an AFOSR MURI.

\end{acknowledgments}

%
\putbib[./FBHighField]
\end{bibunit}
\begin{bibunit}[apsrev4-1]

\newpage\null
\begin{center}{ \bf \large \textsc Supplemental Material}\end{center}
\section*{Experiments}
\begin{figure*}[htbp]
\includegraphics[width=1.5\columnwidth]{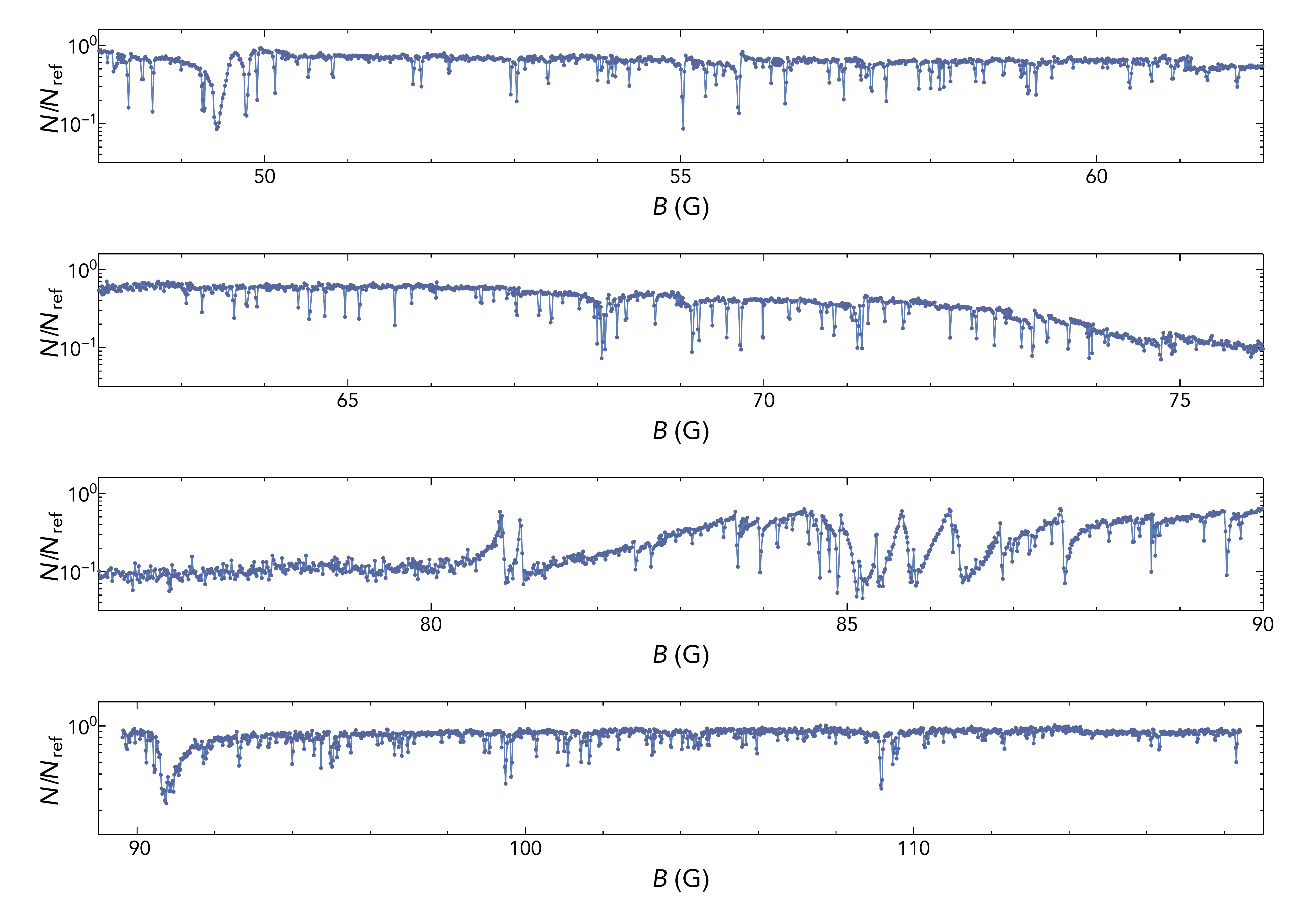}
\caption{Atom-loss spectrum in the surrondings of the broad resonance at $76.9\,\g$. We observe an inversion in lineshape when crossing this resonance (between the two first panels), and then a symmetric lineshape at $109\,\g$ suggesting that the local background scattering length changes sign in this region.}
\label{fig:zoom}
\end{figure*}
\begin{figure*}[htbp]
\includegraphics[width=2\columnwidth]{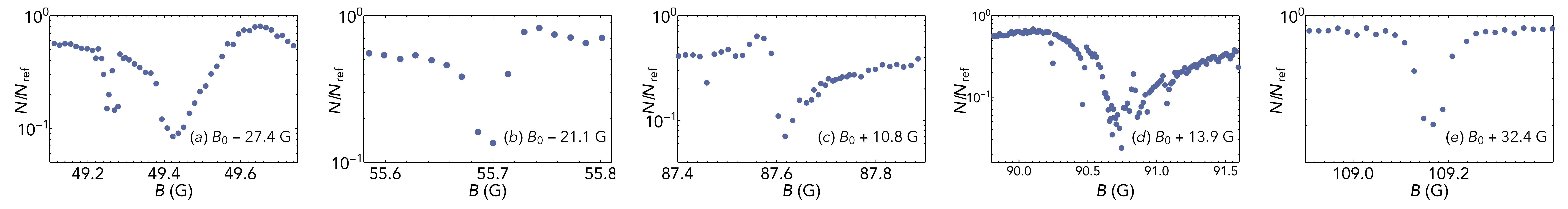}
\caption{Zoom in on narrow resonances, (a) and (b) show resonances with a zero-crossing on their high-field side ($\abgt>0$) which occurs for fields $B<B_0=76.9\,\g$. (c) and (d) are examples with inverted asymmetry occurring for $B>B_0$, implying $\abgt<0$, (note that the resonance in (e) overlaps with several very narrow lines). Finally at $109\,\g$ (e) we observe a symmetric lineshape which corresponds to a $\abgt$ close to zero.}
\label{fig:lineshapes}
\end{figure*}
\begin{figure}[htbp]
\includegraphics[width=.8\columnwidth]{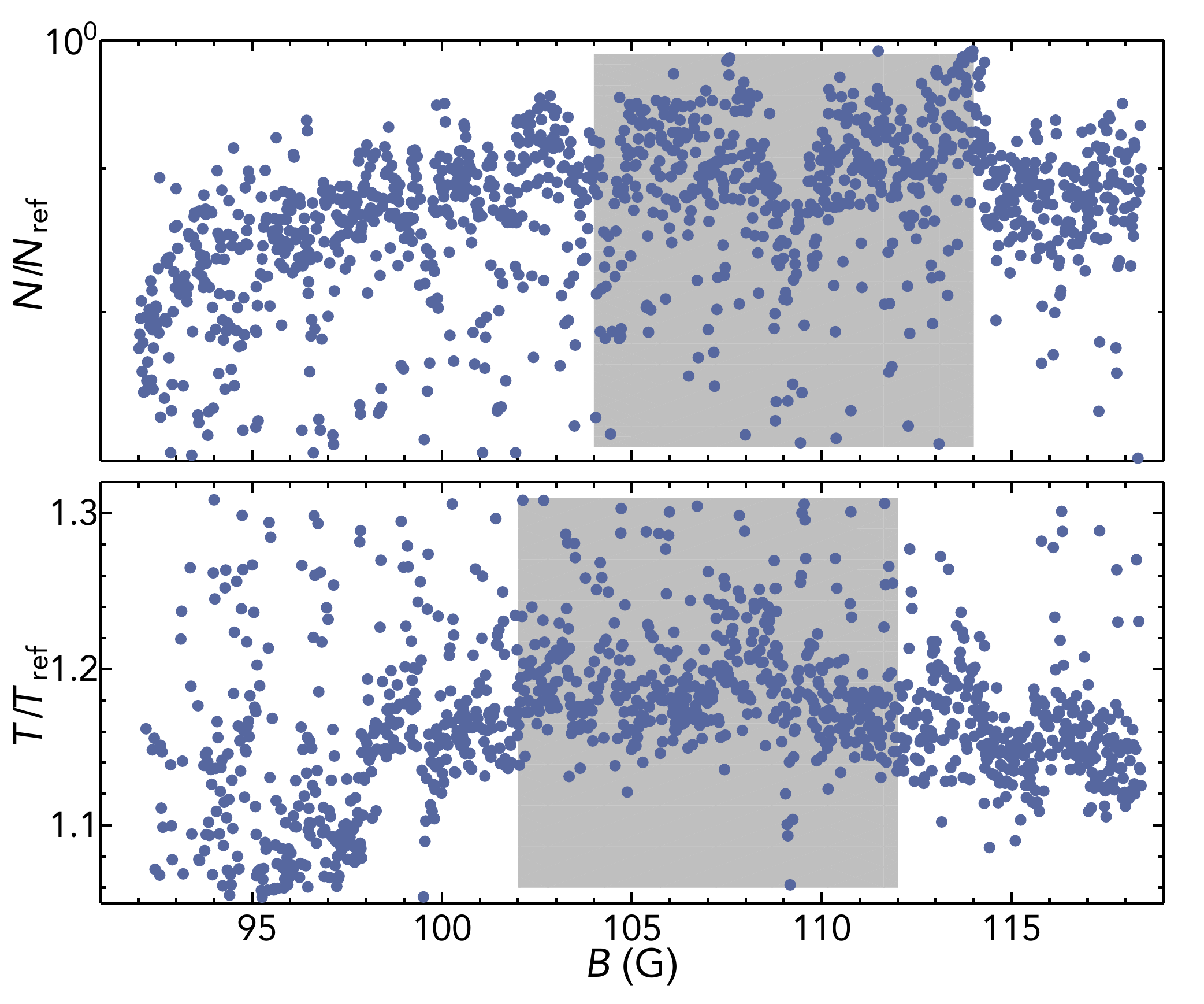}
\caption{Focus on the background atom number and temperature in between narrow resonances after a wait of $500\,\ms$, $N_{\rm ref}$ and $T_{\rm ref}$ are taken after the same wait time at a field close to zero. A maximum is visible on both observables, that we locate for the atom number at $B_{{\rm Max,}N}=109\,(5)\,\g$ and for temperature at $B_{{\rm Max,}T}=107\,(5)\,\g$, shown as grey areas.}
\label{fig:NandT}
\end{figure}

\subsection{Atom-loss spectroscopy}
The early stages of our experiment have been presented in \cite{Maier:2014,Maier:2015}, we detail here the specifics of our atom-loss spectroscopy. The atom-loss spectroscopy consists in recording the number of atom left after a given wait time as a function of magnetic field. On a Feshbach resonance, three body losses are amplified so that it shows as a minimum in final atom number. The parameters used for this technique vary in the different data sets presented throughout the main text. All experiments are performed in a crossed optical dipole trap created by a narrow-line laser with wavelength $1064\,{\rm nm}$. We obtain the trapping potential created by this laser using standard techniques. A forced evaporation is performed in this trap down to the temperature of choice by lowering the beams power. The final power is varied to control the temperature $T$ of the gas. We express the trap depth $U$ through the trap ratio $\eta=U/\kb T$. It can be estimated using the knowledge of the beam waists and the effect of gravity. This is strongly dependent on the knowledge of the trapping potential, furthermore since dysprosium is strongly magnetic, residual magnetic field gradients can change the depth. We then improved our estimations using a statistical analysis of final temperature and atom number in our atom-loss spectroscopy using loss dynamics equations adapted from \cite{Eismann:2015} for finite s-wave scattering length and including universal dipolar scattering. We obtain a knowledge of $\eta$ with an uncertainty of $20\%$.

In figure~[1] of main text we present a spectroscopy performed at a temperature of $2.4\,\muK$. The samples were prepared in the following way: after evaporation at a magnetic field close to $0\,\g$ down to $T\simeq0.6\,\muK$, we recompress the trap by increasing the beams power by a factor of approximately $\times2$ in 100 ms. This has two purposes, first we reduce residual evaporation induced by increased elastic collisions, second we minimize a weakening of the trap confinement caused by a residual field gradient created by our Feshbach coils, and partially to the crossing of many Feshbach resonances. We then increase the bias magnetic field in $15\,\ms$ to the target value. We observe that the turning-on of the magnetic field to values above $60\,\g$ results in a residual heating,  
this heating is partly due to an unintentional change of direction of the magnetic field when ramping to high fields. At the change of sign the cloud is in a zero magnetic field and is slightly depolarized which results in dipolar relaxation and heating when the field is increased. A residual magnetic field gradient caused by our Feshbach coils probably contributes to this heating as well. The initial temperature of the clouds for the atom-loss spectroscopy is $T=2.4\,\muK$, and the starting atom number $N_0=1.8\times10^{5}$. We estimate the trap depth in these conditions to be $\eta_0=\kb T_0/U=7.0\,(1.5)$. The final atom number presented in figure~[1] of main text is recorded after a wait time of $2\,{\rm s}$ at constant magnetic field.

In figure~[2] of main text, the atom-loss spectra presented were obtained at a temperature of $500\,{\rm nK}$ with initial atom number $N_0\simeq1.0\times10^5$, the two data sets are taken at a similar trap depth of $\eta\simeq5$. The wait time is in both cases $500\,\ms$. The data set spanning the lower resonance was taken with a higher mean trapping frequency than the other one which induces a higher initial density, thus this data shows a lower final atom number.

In figure~[3] of main text, we present three different data sets, in different experimental conditions. The data presented in green (color online) is the same as in figure~[1] of main text. The parameters for the three different data sets presented (green, blue and red) are summarized in table~\ref{tab:paramFig3}. We use these parameters to solve numerically the coupled equations (12) and (25) of \cite{Eismann:2015} from which one obtains the expected final atom number for unitarity-limited two and three-body losses in a Bose gas. 
These equations read
\be
\frac{dN}{dt}=\gamma_3\frac{N^3}{T^5}-\gamma_2e^{-\eta}\frac{V_{\rm ev}}{V_{\rm e}}\frac{N^2}{T}\\
\frac{dT}{dt}=\frac T3\left(\frac{5\gamma_3}{3}\frac{N^2}{T^5}-\gamma_2e^{-\eta}\frac{V_{\rm ev}}{V_{\rm e}}(\eta+\tilde\kappa-3)\frac NT\right)
\ee
with
\be
\gamma_2&=&\frac{16\hbar^2\bar\omega^3}{\pi\kb U}\\
\gamma_3&=&\lambda_3\left(\frac{m\bar\omega}{2\sqrt3\pi\kb}\right)\\
\lambda_3&=&T^2\,L_3
\ee
where $L_3$ is given in the main text and the expressions of $\frac{V_{\rm ev}}{V_{\rm e}}$ and $\tilde \kappa$ can be found in ref.~\cite{Eismann:2015}. These equations can then easily be solved numerically remembering that $\eta=T/U$, using the parameters given in table \ref{tab:paramFig3}.

 These predictions are compared to our measurements in figure~[3]. The shaded area represents the range of calculated final atom number we obtain by varying the input parameters of the model by one standard deviation as given in tab.~\ref{tab:paramFig3}. 
\begin{table}[htbp]
\begin{center}
\begin{tabular}{|c|c|c|c|c|c|}
\hline
data set&$N_0\,(10^5)$&$T_0\,({\rm nK})$&$\nu_{x,y,z}\,(\Hz)$&$\eta_0$&$t_{\rm wait}\,(\ms)$\\\hline

green&1.8\,(2)&2400\,(200)&70 \,(10)&7.0\,(1.5)&2000\\
&&&123\,(15)&&\\&&&263\,(25)&&\\\hline

red&0.46\,(2)&500\,(100)&44\,(5)&4.0\,(1.0)&500\\
&&&87 \,(10)&&\\&&&145\,(15)&&\\\hline

blue&1.0\,(2)&450\,(100)&52\,(5)&5.0\,(1.0)&500\\
&&&23\,(5)&&\\&&&260\,(25)&&\\\hline

\end{tabular}
\end{center}
\caption{Experimental conditions for the atom-loss specta presented in figure~[3] of main text. $N_0$ and $T_0$ are the initial atom number and temperature, $\eta_0=U/\kb T_0$ where $U$ is the trap depth. $\nu_{x,y,z}$ refers to the trapping frequencies in the three dimensions, finally $t_{\rm wait}$ is the time we wait before recording the final atom number $N$}

\label{tab:paramFig3}
\end{table}%

\subsection{Localization of the zero-crossing}

We here employ the (a)symmetry of narrow Feshbach resonances lineshapes to extract their local background scattering length, in the surroundings of the $76.9\,\g$ resonance, represented in figure~\ref{fig:zoom}. The sign of the local background scattering length on a narrow resonance $\abgt$ determines the side on which the local zero-crossing takes place, if $\abgt>0$ ($\abgt<0$) it is on the high- (low-) field side of the resonance. The zero-crossing corresponds to an attached minimum in three-body recombination in its vicinity, that appears as a maximum of atom number in the atom-loss spectra. If the resonance is sufficiently broad to extend over several data points, this leads to an asymmetry of its lineshape in the data, as presented in figures~\ref{fig:zoom},~\ref{fig:lineshapes} also visible for instance in spectra from \cite{Frisch:2014}. We are able to extract the sign of $\abgt$ on several resonances at different magnetic fields, and observe the following:
For fields up to the position of the broad resonance at $76.9\,\g$, all the lineshapes of isolated resonances we can observe correspond to a $\abgt>0$ (fig.~\ref{fig:lineshapes}, (a, b)). The symmetry of the lineshapes is inverted when crossing the broad resonance, as seen in fig.\ref{fig:lineshapes} (c,d). Finally, at a field $B=109\,\g$ we observe a resonance with a very symmetric lineshape, which suggest a $\abgt$ close to zero, we conclude that a change of sign of $\abgt$ occurs in this field region. We do not observe sufficiently broad spectra between $110\,\g$ and $120\,\g$. Higher fields are closer to the other broad resonance at $178.7\,\g$. Since it is a broad resonance, it likely provides the background for the narrow ones in its vicinity, and we cannot conservatively infer information about  the $76.9\,\g$ resonance in this region. Our observation of a zero crossing of the background scattering length near $B=109\,\g$ is corroborated by background atom number and temperature measurements. We present in figure~\ref{fig:NandT} the atom number and temperature as a function of magnetic field in an atom-loss measurement. We observe that the background value of these observables (in between narrow resonances) varies slowly with magnetic field and reaches a maximum. From the data in fig.~\ref{fig:NandT}, we locate the maximum in atom number at $B_{{\rm Max,}N}=109\,(5)\,\g$ and temperature at $B_{{\rm Max,}T}=107\,(5)\,\g$. This maximum in atom number and temperature signals a minimum of thermalization due to a minimum of collision rate, which hampers evaporation. The two-body collision rate is related to the elastic scattering cross-section, which is for dysprosium the sum of the constant universal dipolar scattering cross section and that of the short-range s-wave collisions: $\sigma_{\rm tot}=32\pi D^2/45+8\pi a^2$ where $D$ is the dipole length given in main text. Since the dipolar scattering cross section is independent of magnetic field, the observed minimum in collision rate corresponds to a minimum of $a^2$. This minimum appears to be located close to the field at which the local background scattering length is close to zero according to the lineshapes analysis. With this combined analysis, we conclude that indeed the scattering length in between narrow resonances varies slowly and crosses zero at a field $B=108\,(5)\,\g$.

\begin{figure*}[htbp]
\centerline{
\includegraphics[width=.5\columnwidth]{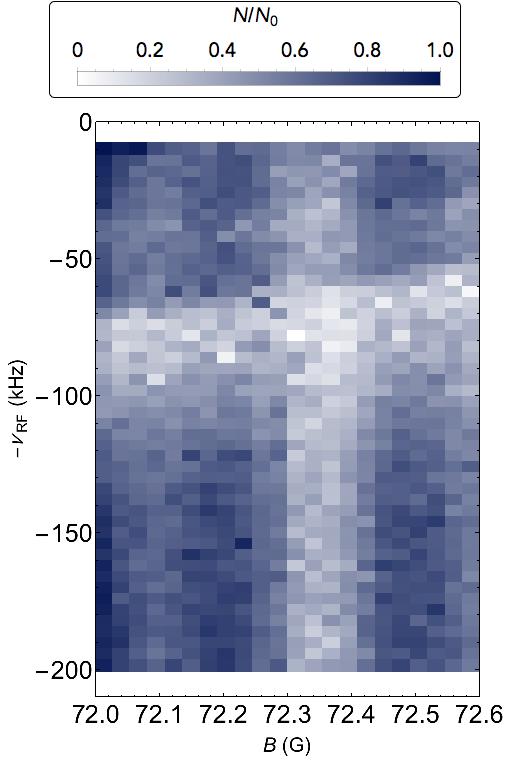}
\includegraphics[width=1\columnwidth]{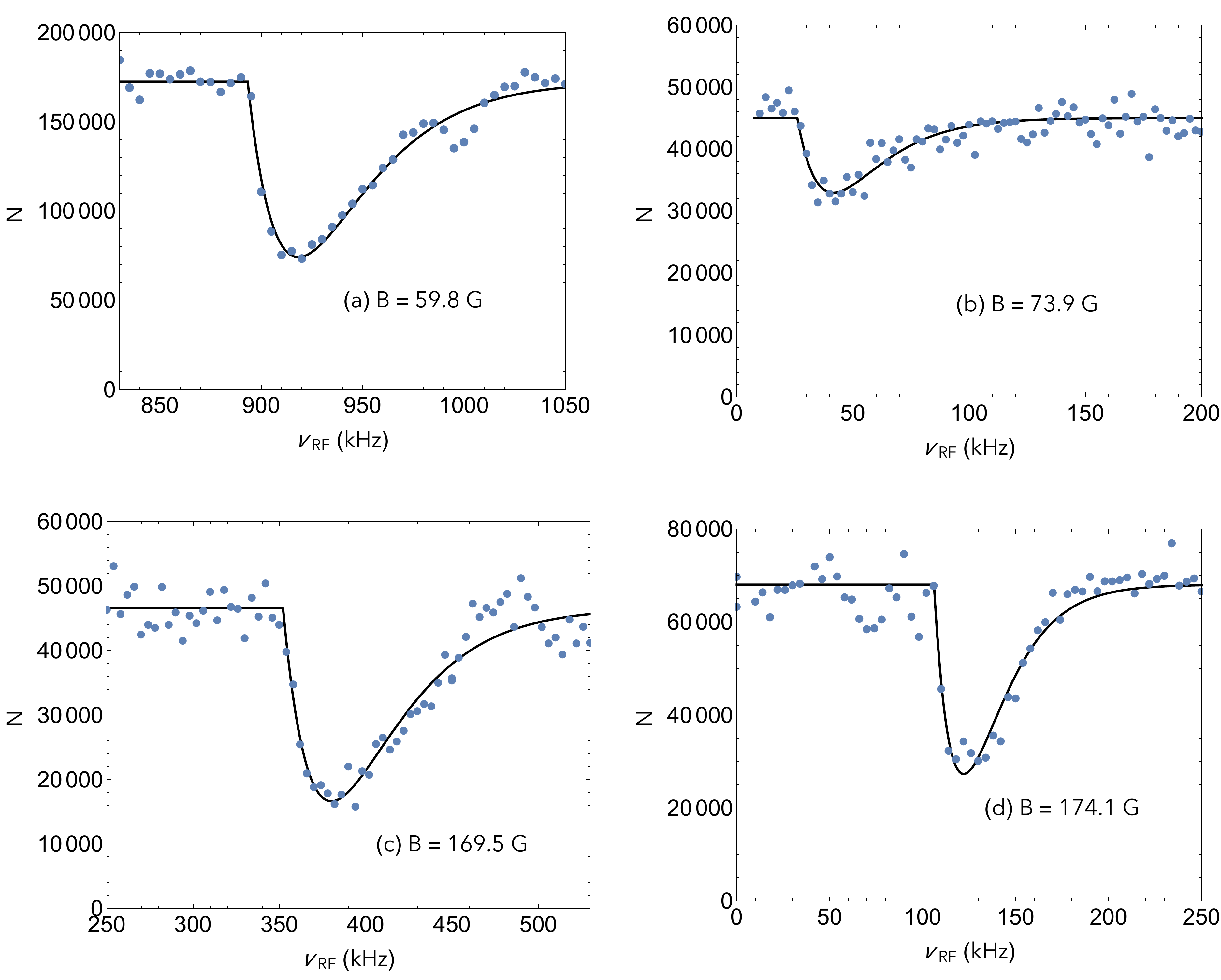}
}
\caption{Magnetic-field Modulation spectroscopy. Left: output of magnetic-field modulation, in the magnetic field range $B\in[72\,\g,\,72.6\g]$. Number of atoms normalized to the number without field modulation $N_0$, as a function of magnetic field and modulation frequency. We observe one narrow, slowly varying feature and one localized very broad feature. The former is the signature of the weakly bound state creating the broad Feshbach resonance at $76.9\,\g$. The latter is created by a bound state responsible for a narrow Feshbach resonance. Right: Examples of magnetic-field modulation spectra from which we obtain the binding energy $\eb$ shown in figure~[2] of main text. The data is fitted by a Maxwell-Boltzmann distribution to extract $\eb$.}
\label{fig:RF}
\end{figure*}

\subsection{Binding energy measurements}

We measure the binding energy of weakly bound dimers via a modulation of the magnetic field around its bias value. This modulation is applied with an additional coil placed near the atom cloud, with a current oscillating at radio-frequencies. We record the number of atoms left after the field modulation as a function of frequency at fixed bias field. The typical amplitude of modulation is $300\,\mg$ during a time of $300\,\ms$. The output of this method presents typically two kinds of spectra, which are exemplified in figure~\ref{fig:RF} (left). We observe either losses in a narrow frequency band, or in a very broad band. We associate very broad spectra to bound states with high energy-vs.-magnetic field slope, meaning that their magnetic moment differs strongly from that of two free atoms. On the opposite, the center frequency of the narrow feature evolves slowly with magnetic field and we associate it with the weakly bound state attached to our broad resonances. We present in figure~\ref{fig:RF} (right) two spectra on each resonance from which we extract the binding energy using a fit with a Maxwell-Boltzmann distribution, solid lines in figure~\ref{fig:RF}. This fit accounts for the broadening of the spectrum by the relative kinetic of two atoms associated. The fit results are typically of good quality given our noise level, the width observed are in qualitative agreement with our temperature. A residual heating of about $20\%$ increase in $T$ is observed when dimers are associated, which we attribute to energy released by the relaxation of shallow dimers to deeply bound states. This dynamics surely modifies our spectra, but since the binding energy is given by the activation energy, meaning the lower bound of the line, its measurement is independent of details of the lineshape. To account for possible systematic errors, we still put conservative error bars on our data. Finally, in this figure one can observe a weak association feature on the left panel at low frequencies, this is the sub-harmonic of the main association line.
\begin{figure}[htbp]
\centerline{
\includegraphics[width=\columnwidth]{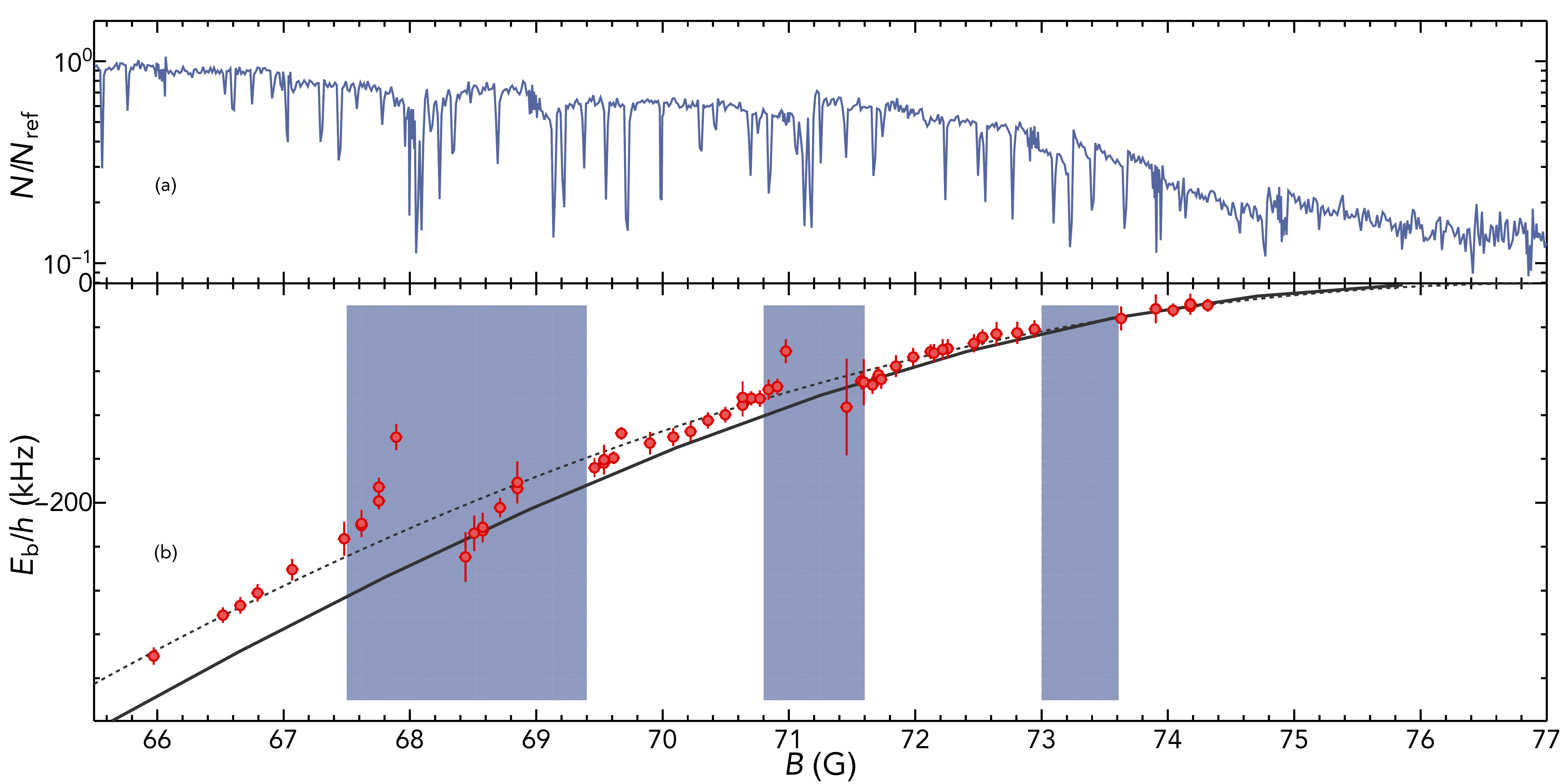}}
\caption{Zoom on the $77\,\g$ resonance, same as fig.~[2] of main text. We outline here in blue the regions where we observe strong avoided crossings in the binding energy as well as stronger-than-usual narrow Feshbach resonances. In these regions the scattering length is unknown. Outside of them resonances are very narrow and modify $a$ only very locally.}
\label{fig:AvoidedCrossings}
\end{figure}

\subsection{Avoided crossing regions}

In the binding energy data we observe avoided crossings that modify locally the halo-dimer binding energy. They result from a stronger-than-usual coupling between the halo dimer and another state. These influence also the scattering length and we thus also observe their effect in the atom-loss spectrum. In these regions the scattering length obtained  in the main text does not hold true, outside of them all resonances are very narrow ($\lesssim20\,\mg$ in the atom-loss spectrum) and perturb the scattering length only very locally. In figure \ref{fig:AvoidedCrossings} we outline in blue the regions where the extracted $a(B)$ does not hold.

\newpage\null
\section*{Theory}

\subsection{Long-range interactions of Dy atoms}
Collisions of ultracold lanthanide atoms such as Dy are very challenging to describe theoretically due to the open $4f$ electronic shell and the resulting large electronic angular orbital momentum. Another difference between Dy and alkali metal atoms is the strong magnetic dipole of $9.93\mu_\mathrm{B}$. The anisotropic dipolar and van der Waals interactions are so strong that ``partial wave'' $L$ is not a good quantum number for the eigenstates of Dy$_2$ molecules, but many different electronic symmetries and partial waves are mixed by the strong anisotropy.  Furthermore, the density of states is very high, corresponding to a mean spacing between narrow resonances on the order of 0.3G for the bosonic species of Dy~\cite{Baumann:2014,Maier:2015} as well as the similar species Er~\cite{Frisch:2014}.  In fact, the series of very narrow resonances observed by Frisch {\it et al.}~\cite{Frisch:2014} is known to have a mean spacing explained by the Wigner-Dyson distribution predicted by random matrix theory, implying a ``chaotic'' set of levels that do not have any assignable set of quantum numbers.  Therefore, the appearance of the broad identifiable $s$-wave bound states in the present experiment is somewhat of a surprise, since this well-characterized universal state is quite different in character from the set of ``chaotic'' levels that make the dense set of narrow resonances.   The long-range $s$-wave ``halo'' state effectively decouples from the ``sea'' of chaotic levels and generates a broad resonance that spans a relatively large tuning range of $B$. In this case, for the magnetic field regions of our current interest, the impact of narrow resonances is negligible outside very small regions around their locations. Each wide resonance provides a ``local'' background for nearby narrow ones, as illustrated for alkali atoms in Refs.~\cite{Berninger:2013,Jachymski:2013}.

The experiment prepares the Dy atoms in the maximal projection state with angular momentum $j=8$ and projection  $m=-8$. The dipolar physics of two Dy atoms can be set up in different basis sets.  Kotochigova and Petrov~\cite{Kotochigova:2011b} set up the usual body-frame Born-Oppenheimer electronic states, whereas Bohn {\it et al.}~\cite{Bohn:2009} set up a space-frame asymptotic basis in a partial wave expansion.  Since we are primarily concerned here with the long-range properties of universal $s$-wave states, we will use the latter to describe the collisions of two Dy atoms in the same Zeeman sublevel.  This requires a single $s$-wave entrance channel, although we will need to consider its off-diagonal coupling to the $d$-wave channels.   The many other electronic and partial wave channels only come into play by generating a dense set of threshold crossing levels with very different magnetic moments that are weakly coupled to the $s$-wave halo state.

The basic understanding of the $s$-wave halo state requires only a knowledge of the long-range potential, which is comprised of one or more terms of the form $C_p/R^p$.  Here $p$ can be 6, 3, or 4 for the respective van der Waals, dipolar, and non-adiabatic contributions~\cite{Bohn:2009}.  There are several definitions in the literature for the scale length  of a $C_p/R^p$ long-range potential, and we adopt here the definition of Gao~\cite{Gao:2008}
\begin{equation}
\label{eq:Gao_def}
 R_p = \left ( \frac{2\mu C_p}{\hbar^2} \right )^{\frac{1}{p-2}}\,,\,\,\,\,E_p = \frac{\hbar^2}{2\mu R_p^2}\,.
\end{equation}
Note that the lengths $R_\mathrm{vdW}$ and $\bar{a}$ defined for the $p=6$ van der Waals potential in Ref.~\cite{Chin:2008} are respectively $\frac12 R_6$ and $0.477989 R_6$.  Note also that the dipole length $D$ defined by Bohn {\it et al.}~\cite{Bohn:2009} is $\frac12 R_3$.   These lengths define corresponding energies $E_p$ as in Eq.~\ref{eq:Gao_def}.

Bohn {\it et al.}~\cite{Bohn:2009} consider the collisions of two dipoles with their orientation held fixed in space by an external field.  In our case, each atom maintains its space projection $m_j = -8$ when the distance between two colliding atoms is large enough that the Zeeman splitting due to the laboratory $B$ field is large compared to the diagonal and off-diagonal potential terms due to the van der Waals and dipolar interactions.  This is the case at distances larger than the shortest length scale $R_6$ needed or analyzing the $s$-wave halo state.  Thus, each atom stays in the same $|j,m\rangle=|8,-8\rangle$projection state and the asymptotic basis set is the set of all states $|8,-8\rangle_1|8,-8\rangle_2|LM_L\rangle$, which we need to label only by the partial wave quantum numbers $L,M_L$.  For identical bosons we need only even $L$ values, and only partial waves with $M_L =0$ are in the same Hamiltonian that contains the $s$-wave, since the total projection quantum number $-16$ is conserved.    Treating the short-range states and the dense set of chaotic eigenstates would require including the full basis set, $|jm_1\rangle_1|jm_2\rangle_2|LM_L\rangle$ for all separated atom combinations and the mixing of many partial waves~\cite{Frisch:2014}, but we can ignore this complexity in estimating the effect of $s$-wave universality, and incorporate the effect of such complex states as causing the narrow avoided crossings with the universal $s$-wave halo state in the near-threshold region.

The asymptotic space-frame physics is described by Eqs. (2)-(5) of Bohn {\it et al.}~\cite{Bohn:2009}, where we will estimate the effect of dipolar interactions using only the $|LM_L\rangle=$ $|0,0\rangle$ and $|2,0\rangle$ terms in the Hamiltonian, described by the coupling coefficients (in the notation of Ref.~\cite{Bohn:2009}) $C^{(0)}_{00}=0$, $C^{(0)}_{02}=1/\sqrt{5}$, and $C^{(0)}_{22}=2/7$.  Thus, the diagonal dipole interaction $-C_3/R^3$vanishes for the $s$-wave, and is $C_3=2C^{(0)}_{22}d^2$ for the diagonal $L,M_L=2,0$ $d$-wave component (the other $d$-wave components have $C^{(\pm 1)}_{02}=1/7$ and $C^{(\pm 2)}_{02}=-2/7$ and contribute to universal dipolar scattering, but do not couple to the $s$-wave).  Since for Dy $d=9.93 \mu_B$ and $\mu_B = \alpha/2$ atomic units, where $\alpha$ is the fine structure constant, we find that $d^2=0.00131$ atomic units (1 atomic unit $= E_\mathrm{h}\mathrm{a}_0^3$, where $E_\mathrm{h}$ is one Hartree and a$_0$ is the Bohr radius).  The Bohn scaled unit of length $D =$ 196a$_0$ and the Gao one is $r_3=$ 392 a$_0$ for $C_3=d^2$.  

Although dipolar interactions vanish in the diagonal $s$-wave potential, Bohn {\it et al}~\cite{Bohn:2009} show how a simple perturbation estimate shows that the off-diagonal coupling between the $s$- and $d$-wave $M_L=0$ components give an effective adiabatic potential that varies at long-range as $-C_4/R^4$.  This is because the splitting between these components at long-range is given by the centrifugal potential, $6\hbar^2/(2\mu R^2)$, which is much larger than the off-diagonal dipolar term $(2/\sqrt{5})d^2/R^3$.  The long-range adiabatic coefficient is $C_4=(4/15)d^2D$ $=$ 0.0687 atomic units $E_\mathrm{h}\mathrm{a}_0^4$.  This gives a Gao scale length of $R_4=$ 143 a$_0$.  This is comparable to the scale length of the van der Waals potential, for which $C_6=$ 1890 au~\cite{Kotochigova:2011b}, $R_6=$ 154 a$_0$, and $E_6/h=$ 0.9265 MHz.

\subsection{Overlapping Feshbach resonances}

Many alkali metal atom systems are known to have broad $s$-wave resonances on which are superimposed one or more much narrower resonances of higher partial wave bound state character; for example, Ref.~\cite{Berninger:2013} for Cs $+$ Cs gives an example with an overall very large $a_{bg}$ and Ref.~\cite{Cho:2013} for Rb $+$ Cs gives an example with a quite small $a_{bg}$.  Reference~\cite{Jachymski:2013} developed an analytic formalism based on multichannel quantum defect theory (mqdt) to characterize such sets of overlapping resonances.  The mqdt treatment assumes the following: (1) one $s$-wave entrance channel with ``background'' scattering length $a_{bg}$ and $p=6$ length $R_6$, (2) a set of $N >1$ closed channels each with an uncoupled ``bare state'' with  crossing positions $B_{i}$ ramping according to magnetic moment differences $\delta \mu_i$ from the entrance channel, and (3) coupling between the entrance channel and closed channel $i$ described by a dimensionless coupling matrix elements $y_i$, or equivalently, a set of ``bare'' resonance ``widths'' $\Delta_i$.  The result of the analysis is that the scattering length with a set of $N$ poles can be put in the following form:
\begin{equation}
\label{eq:a_overlap}
a(B)=a_{bg}\prod_{i=1}^N{\left(1-\frac{\tilde{\Delta}_i}{B-B^{res}_i}\right)}.
\end{equation}
Here $a_{bg}$ provides a ``global'' background for all the resonances, and the  ``widths'' $\tilde{\Delta}_i$ and resonance pole positions $B^{res}_i$ are generally functions of all the ``bare'' widths $\Delta_i$ and crossing positions $B_i$.

One consequence of the general formula is that near any particular pole $i$ we can write
\begin{equation}
\label{eq:a_local}
 a(B) = \tilde{a}_{bg,i} \left ( 1 - \frac{\tilde{\Delta}_i}{B-B^{res}_i} \right ) \, ,
\end{equation}
where the ``local'' background scattering length is
\begin{equation}
\tilde{a}_{bg,i} =a_{bg}\prod_{j\ne i}^N{\left(1-\frac{\tilde{\Delta}_j}{B^{res}_i-B^{res}_j}\right)} \,.
\end{equation}
Here only the product $\tilde{a}_{bg,i} \tilde{\Delta}_i$ is well-defined, whereas the individual terms can depend on the number of resonances and $B$ field range included in the model.  Thus, the ``local'' pole strength of a resonance is well-defined, and certainly there are cases where a broad $s$-wave resonance can be characterized by a single $s_6$ parameter as $B$ is tuned across several narrow resonances.  A good example is provided by the pair of $s$- and $g$-wave resonances of two Cs atoms with respective positions and pole strengths of $(B_i, s_{res,i}) = $ $(548.44\mathrm{G}, 160)$ and $(554.06\mathrm{G}, 1.6)$~\cite{Jachymski:2013}.  Accounting for the overlapping properties of these two resonances is essential in characterizing the three-body physics in the 550 G region of Cs~\cite{Wang:2014}. 

\begin{figure}
\centering
\includegraphics[width=0.4\textwidth]{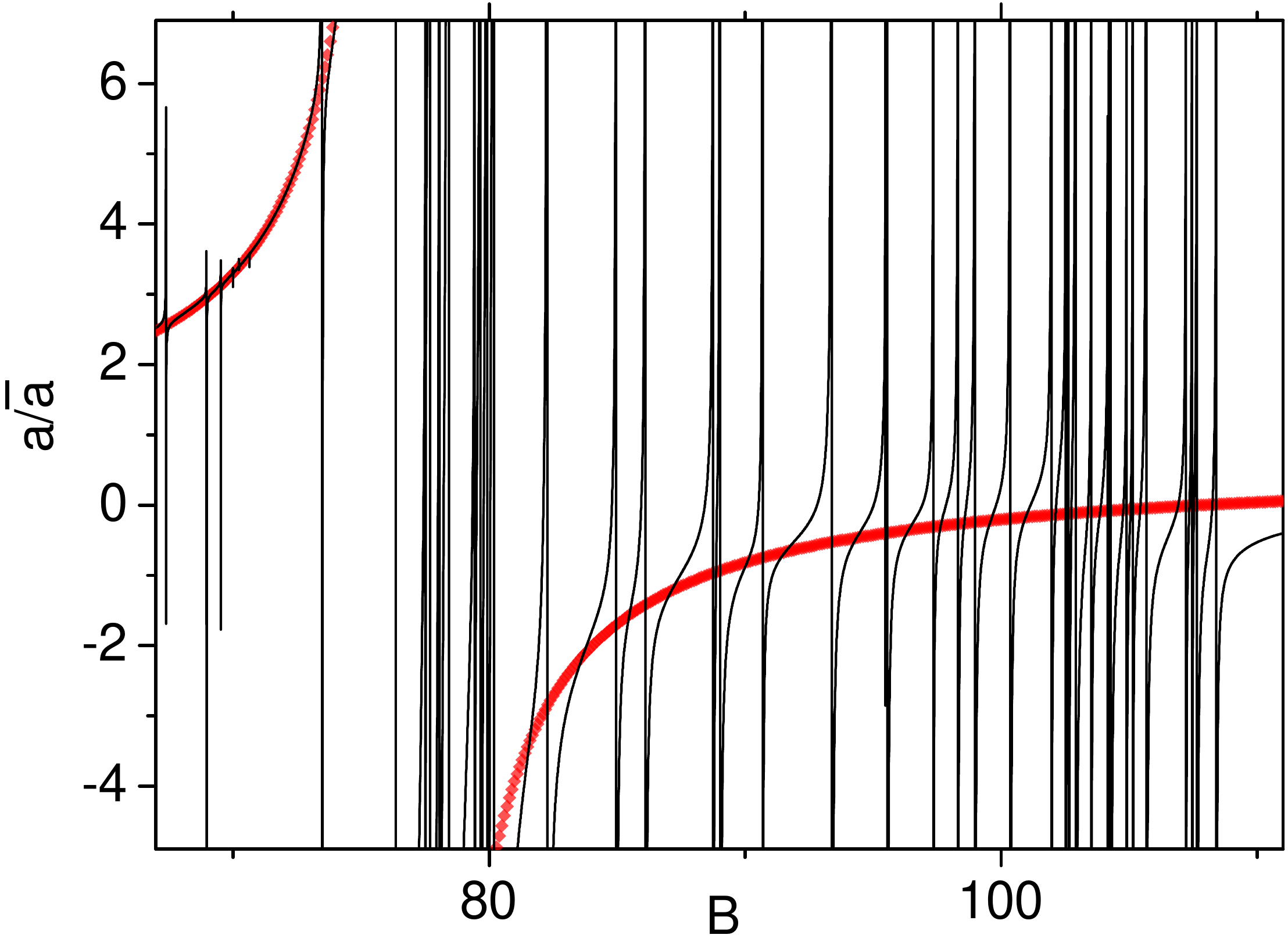}
\includegraphics[width=0.4\textwidth]{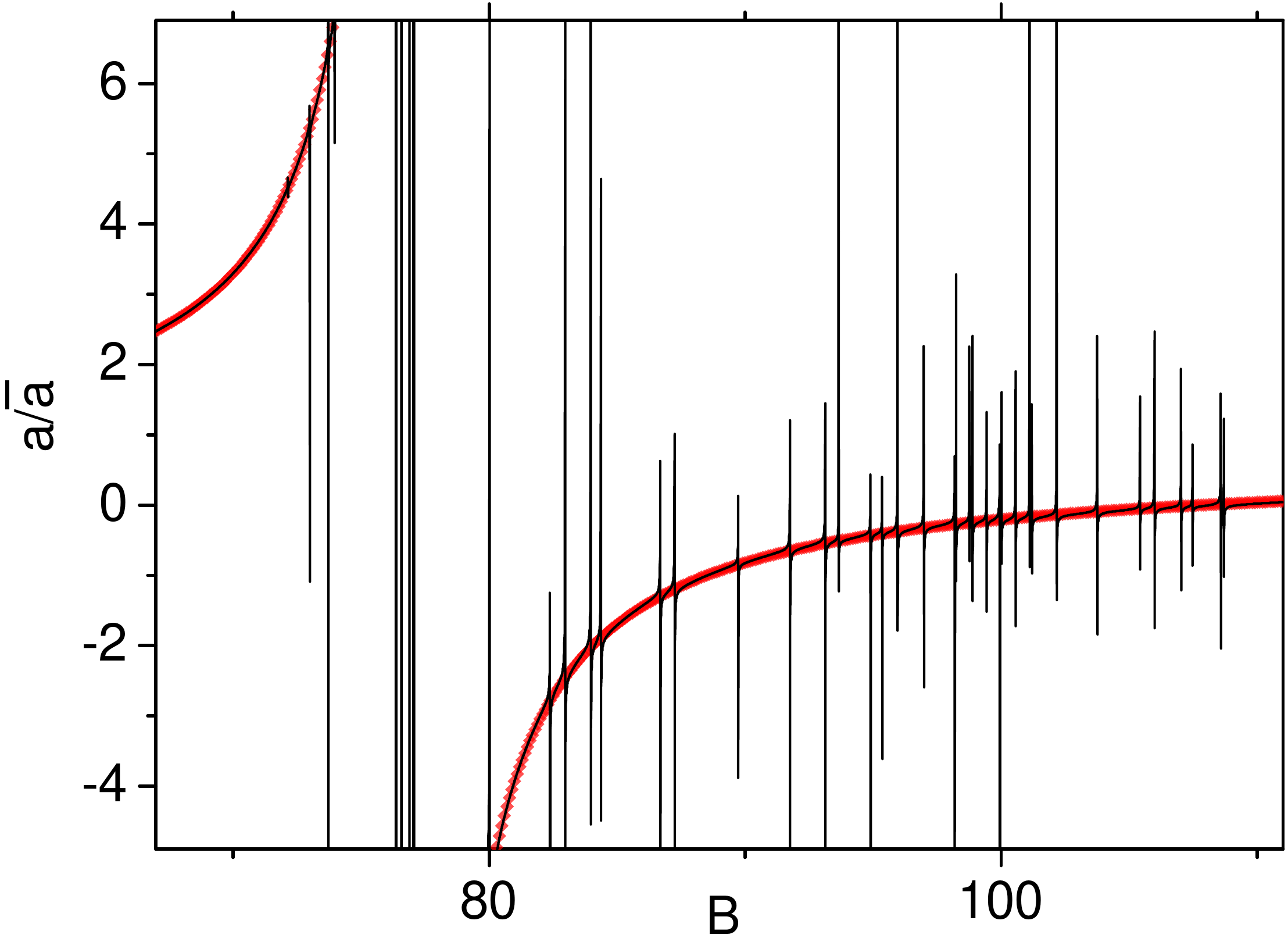}
\caption{\label{overlapping}Scattering length for the case of a single resonance similar to the 77G one observed in the experiment overlapping with 50 much narrower ones (upper panel: mean width of narrow resonances is 0.125 G, lower panel: mean $\Delta$ equals 0.004 G). The red thick curve shows the background induced by the wide resonance.}
\end{figure}

Figure~\ref{overlapping} gives an example of how the scattering length can look within this treatment for the case of a single very broad resonance overlapping with many narrow ones.  In this example with a toy model, a random distribution of narrow resonances was taken with two different values assumed for the mean width.  The broad resonance provides the ``local'' background $\tilde{a}_{bg,i}$ for the narrow resonances.  The range over which a narrow resonance perturbs the underlying background due to the strong resonance depends on the width of the narrow one.  Figure~\ref{fig:AvoidedCrossings} indicates regions where such perturbations are evident in the binding energy curve.  It will be important for future theoretical and experimental work to investigate the pole strengths of such narrow resonances and assess their effect.   Reference~\cite{Nicholson:2015} describes how it is not valid to extrapolate the properties of a broad resonance across a range on the order of the spacing between different vibrational levels of the same channel, although it is valid to extrapolate its properties across a set of resonances in different spin channels.

\subsection{Universal long-range theory for hybrid interaction potential}

The scattering length $a(B)$ and the energy $E(B)$ of the last $s$-wave bound state near an isolated magnetically tunable near-threshold Feshbach resonance is given by~\cite{Chin:2008}:
\begin{equation}
\label{eq:a+E}
 a(B) = a_{bg} \left ( 1 - \frac{\Delta}{B-B_0} \right ) \, , \,\,\,\, E(B) = -\frac{\hbar^2}{2\mu a(B)^2} \,
\end{equation}
where $a_{bg}$ is the background scattering length of the $s$-wave entrance channel in the absence of the resonance, $B$ is magnetic field strength, $B_0$ is the pole position of the resonance as collision energy $E \to 0$, and $\Delta$ represents the width of the resonance.    The ``bare'' closed channel level that makes the resonance ramps linearly with $B$ as $\delta \mu (B-B_c)$, where $\delta \mu$ is the magnetic moment difference between the closed channel state and the two separated atoms.  The actual pole position $B_0$ is shifted from the ``bare'' crossing location $B_c$ by a shift that can be as large as $\Delta$, depending on $a_{bg}$~\cite{Chin:2008}.  While the formula for $a(B)$ applies over a wide range of $B$, the ``universal'' binding energy formula for $E(B)$ only applies very close to threshold where $a \gg R_6$, where $R_6$ is the scale length of the van der Waals potential.  

The ``pole strength'' of resonance term in Eq.~\ref{eq:a+E} is characterized by the product $a_{bg}\Delta$ and more generally by a dimensionless parameter $s_6 = (a_{bg}/R_6) (\delta \mu \Delta/E_6)$, or equivalently, by the $\sres=0.477989 s_6$ parameter used in Ref.~\cite{Chin:2008} to classify resonances (see also Ref.~\cite{Nicholson:2015}).  Strong or ``broad'' resonances have $\sres \gg 1$ and weak or ``narrow'' resonances have $\sres \ll 1$.  The former have universal $s$-wave bound states as $B$ is tuned over a large fraction of the resonance width $\Delta$ whereas the latter are universal only over a small fraction of their (generally small) $\Delta$ near the pole.  Reference~\cite{Julienne:2014} studied in detail the departure from universality of the binding energy for broad and narrow resonances in the $^6$Li and $^7$Li systems.

\begin{figure}
	 \includegraphics[width=1\columnwidth]{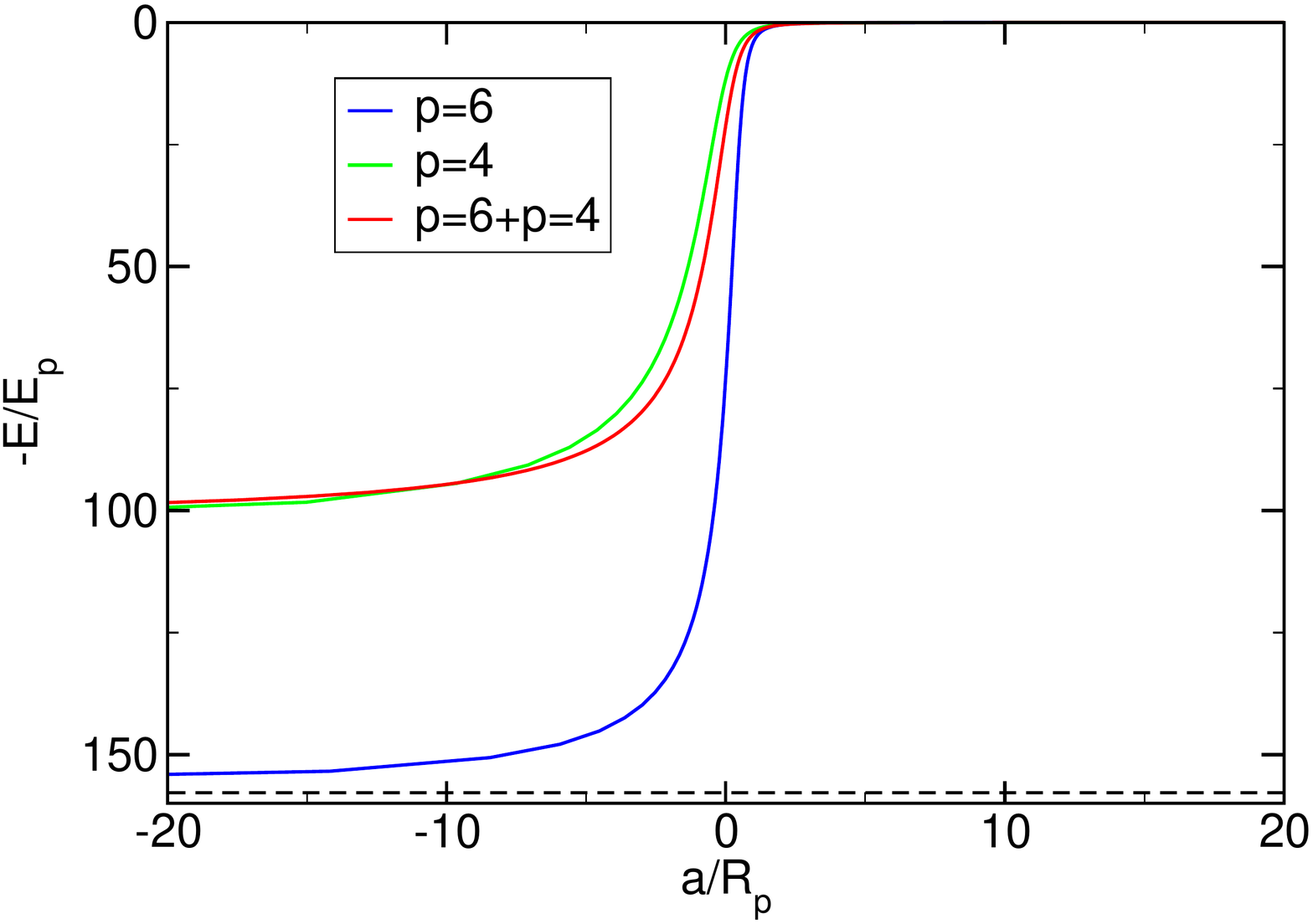}
	 \includegraphics[width=1\columnwidth]{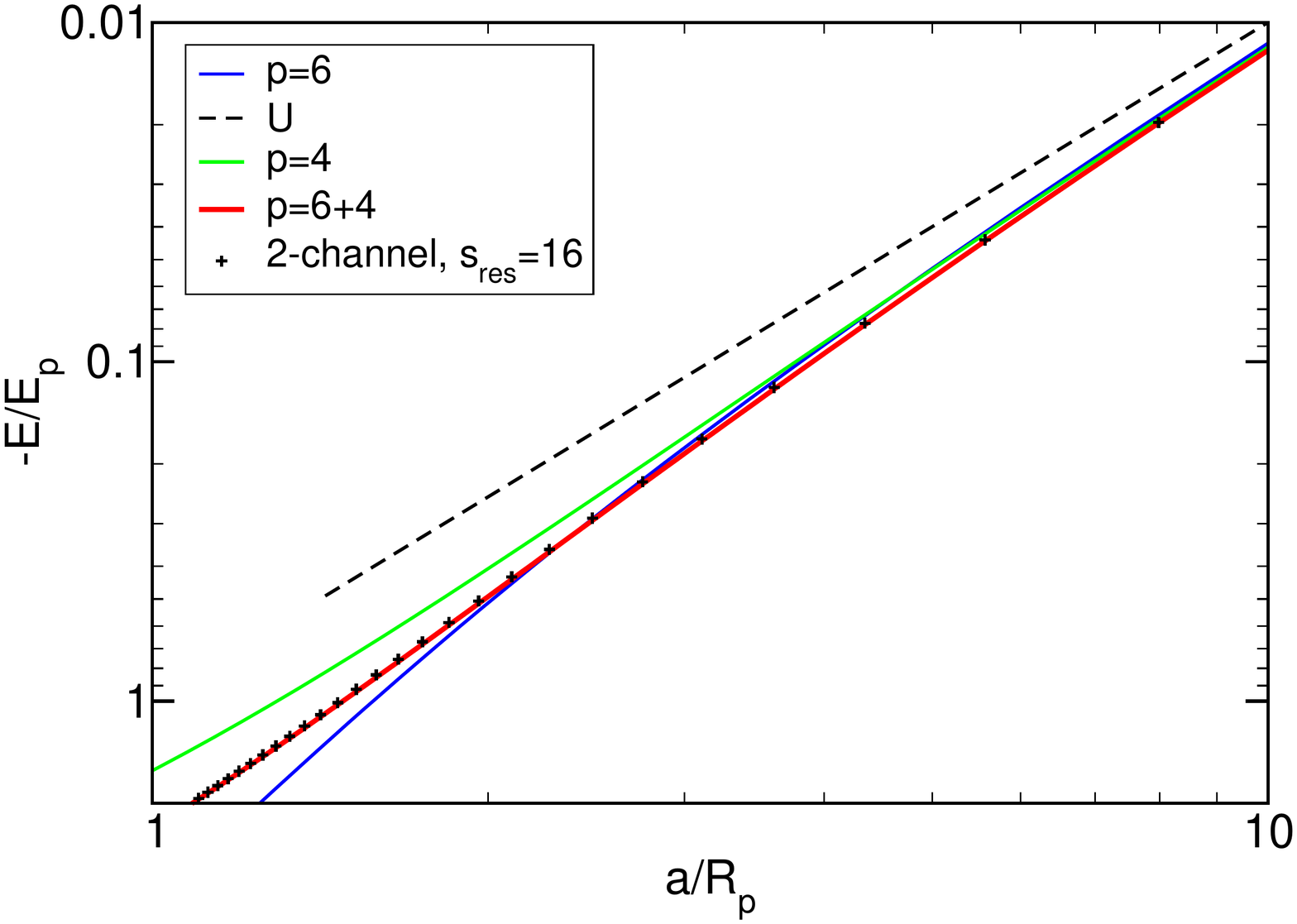}
	 \caption{(Upper Panel) Single channel binding energy curves $E/E_p$ versus scattering length $a/R_p$ for $p=$ 4, 6, and for a hybrid potential with both $p=$ 4 and 6 terms to simulate the $s$-wave collision of two weakly dipolar Dy atoms.  The $C_4/C_6$ ratio for the latter is chosen to be the value from the perturbation analysis for two Dy atoms (see text).  The hybrid case is scaled by $E_6$ and $R_6$ from the $p=6$ term.  The dashed line shows the bottom of the ``bin'' for the last bound state of the van der Waals potential at $E/E_6=-157.8$.  (Lower Panel) Same as above but on a log scale near threshold.  The dashed line shows the universal quadratic relation in Eq.~\ref{eq:a+E}, to which the different curves all tend near threshold when $a/R_p \gtrsim 10$.  The departures from quadratic universality follow similar curves for the $p=4$, 6, and hybrid potentials for $a/R_p \gtrsim 3$, but these cases show variation from one another at smaller $a$.  The crosses show the coupled channels curve from the 2-channel calculation with $\sres=16$.  For Dy, $E_6/h=$ 0.926 MHz and $R_6=$ 154 a$_0$.
	  }
	 \label{fig:E_vs_a}
\end{figure}

Using the Gao scaled units, $r=R/R_p$ and $\epsilon = E/E_p$, the single channel Schr{\"o}dinger equation for a power law $-C_p/R^p$ potential becomes 
\begin{equation}
\label{eq:Schrodinger}
 \frac{d^2\psi}{dr^2} +\left ( \epsilon + \frac{1}{r^p} - \frac{L(L+1)}{r^2} -v_{sr}(r) \right )\psi =  0 \,,
\end{equation}
where $v_{sr}(r$ represents the short-range contribution.  For numerical purposes we can simply represent this short-range part as a Lennard-Jones $p-2p$ potential, as $v_{sr}(R) = +\sigma^p/r^{2p}$, where the short-range parameter $\sigma$ is adjusted to select the number of bound states in the potential and the scattering length.  When the number of bound states varies from $n$ to $n+1$ as $\sigma$ is varied over a range $\sigma_n$ to $\sigma_{n+1}$, the scattering length of the power law potential can be varied from $-\infty$ to $+\infty$.  Similarly, the binding energy of the last bound state can be varied from its largest possible value at the bottom of the last ``bin'' in which an $s$-wave bound state must be found to zero energy when the scattering length approaches $+\infty$. Such a calculation gives the $E(a)$ dependence for a single channel and describes open-channel-dominated resonances with $\sres\gg 1$. In this sense, open-channel-domianted resonances are all universal, although they show potential-dependent deviations from the qudratic universality described by~\eqref{eq:a+E}. 

Figure~\ref{fig:E_vs_a} shows the $\sres\gg 1$ binding energy curves $E(a/R_p)/E_p$ for $p=6$ and $p=4$ in scaled units. These curves are ``universal'' in that they apply to any species and are independent of reduced mass and the $C_n$ coefficient by using scaled units.  The bottom of the ``bin'' for the last bound state is at $-158E_6$ and $-105E_4$ for the respective $p=$ 6 and 4 potentials.  The Figure also shows the $E(a)$ curve, scaled by $E_6$ and $a_6$, for the hybrid adiabatic $s$-wave potential for Dy with $C_6=1890$ au and $C_4=0.0686$ au, as described previously.  In this case, $E_6/h=$ 0.926 MHz and $R_6=$ 154 a$_0$.

\begin{figure}
	 \includegraphics[width=1\columnwidth]{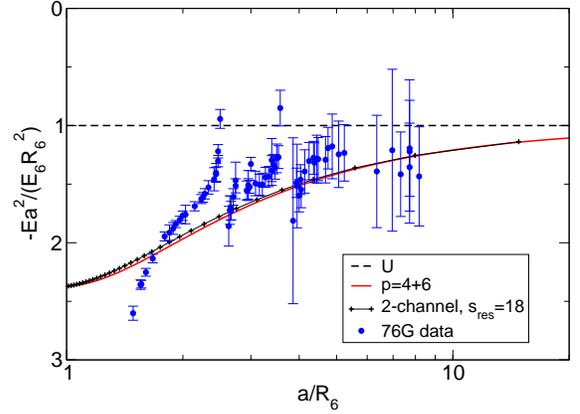}
 \caption{Plot of the information in the lower panel of Fig.~\ref{fig:E_vs_a} in the form $Ea^2/(E_6 R_6^2)$, which approaches unity for the universal relation in Eq.~\ref{eq:a+E}.  The clear departure from the universal scaling due to the finite range of the potential is evident in the range of $a/R_6 <10$, which is the range sampled in the Dy experiment, indicated by the data points, where each point at $B=B_i$ is plotted using the $a(B_i)$ mapping given by the coupled channels calculation.  The departure of the data from the predictions of the hybrid potential model at smaller $a/R_6$ is likely due to the breakdown of our over-simplified 2-channel model as binding energy becomes larger and $a$ smaller.
 }
	 \label{fig:Ea2_vs_a}
\end{figure}

The lower panel of Fig.~\ref{fig:E_vs_a} shows an expanded view on a log scale of the region near threshold, where the conventional ``universal'' quadratic binding energy curve applies in the limit $a \to \infty$ based on the approximation in Eq.~\ref{eq:a+E} that $E/E_p = (R_p/a)^2$.   Figure~\ref{fig:Ea2_vs_a} shows the same information plotted as $E a^2/(E_6 R_6^2)$ versus $a/R_6$, which approached unity as $a \to \infty$ (see Ref.~\cite{Julienne:2014}).  The actual binding energy curves are very close to the universal form when the scattering length is $10R_p$ or larger, but begin to show noticeable variation from the universal curve when the scattering length is smaller.   The actual binding energy differs by a factor of 2 from the universal curve when the scattering length is as small as $2R_6$.  Therefore, a much better mapping of measured binding energy to scattering length will be obtained if the actual scaled $E(a)$ curve is used instead of the universal one.

\subsection{Coupled channels model of broad Feshbach resonances}
A full coupled channel model describing the properties of the wide resonant features observed in the experiment is beyond our reach. Instead, we turn to a simplified two-channel description, neglecting the impact of the narrow resonances, which in this case seems to be a good assumption based on the experimental data, namely, that a universal $s$-wave halo state emerges from the sea of chaotic levels due to the ramping with $B$ of a particular approximate eigenstate to which the entrance channel is strongly coupled.  We construct a 2-channel model using the method described in Refs.~\cite{Mies:2000,Nygaard:2006,Wang:2014} for reducing the full coupled channels model with many resonances to an effective 2-channel one based on potentials having the right long-range form, but with a reduced number of bound states.  We assume a Hamiltonian matrix of the form:
\begin{equation}
\label{eqn:CC_V}
\bf{V} = \left( \begin{array}{cc}
V_g(R) & A_{res}e^{-\beta R}  \\
 A_{res}e^{-\beta R} &  V_c(R)+E_{res}
 \end{array} \right ) \, ,
\end{equation}
where $V_g(R)$ describes the hybrid ground state adiabatic potential with the 6-12 LJ form plus the $p=4$ term, the off-diagonal term describes the coupling between the entrance and the closed channel with potential $V_c(R)+E_{res}$ and $E_{res}>0$ is the asymptotic energy of the closed channel; we assume $V_g$ and $V_c$ both vanish as $R \to \infty$.  The inner wall of $V_g$ is varied to select the scattering length and the number of bound states in the entrance channel.  We chose to set $a_{bg}=102$ a$_0$ and the number of bound states to be 13.  As in Refs.~\cite{Mies:2000,Nygaard:2006}, we take $V_c$ to be the same potential as $V_g$.  The $-2$ vibrational level with a binding energy of 230.507 MHz was selected to represent the Feshbach level; thus, the ``bare'' Feshbach level will be at the $E=0$ threshold of the entrance channel when $E_{res}/h=E_{res}^c/h=$ 230.507 MHz.   Varying $E_{res}$ thus allows us to ``ramp'' the bare Feshbach level across threshold and tune the scattering length of the coupled system, which will generally have a pole at a shifted position $E_{res}^0 \neq E_{res}^c$.  If we assume $E_{res}$ is a linear function of $B$ with slope $\delta \mu$, we can select the mapping of experimental $B$ to the Hamiltonian model parameterized by $E_{res}$:
\begin{equation}
\label{eq:Eres_to_B}
 E_{res} = E_{res}^c + \delta \mu (B - B_c)= E_{res}^0 + \delta \mu (B - B_0)
 \end{equation}
where $B_c$ is the field where the bare Feshbach level crosses threshold.  The pole of the scattering length generally occurs at a different field $B_0$, with the shift from $B_c$ being described in Ref.~\cite{Chin:2008}.

The inter-channel coupling $V_{gc}(R)$is characterized by the off-diagonal term proportional to $A_{res}$, which is selected to give the resonance ``pole strength,'' or $\sres$ parameter; here we take $\beta=1 $ a$_0^{-1}$.  The energy-dependent width $\hbar\Gamma(E)$ of the resonance for collision energy $E>0$ is (see Eq. 36 of Ref.~\cite{Chin:2008})
\begin{equation}
\label{eq:width}
 \hbar\Gamma(E) = 2 \pi | \langle b | V_{gc}(R) |E \rangle |^2 = 2(k\bar{a})(\bar{E} \sres)
\end{equation}
where $|b\rangle$ and $|E\rangle$ are the wave functions of the respective bound state and energy-normalized $s$-wave scattering state, $\hbar k$ is the relative collision momentum; equivalently, we could define $\hbar\Gamma(E)=2k a_6 E_6 s_6$.

Given that we have fixed a value of $a_{bg}$, which we here take to be ``nominal'' at a magnitude order of $100$ a$_0$, the free parameters in the model are $E_{res}$ for resonant tuning and the strength of off-diagonal coupling $A_{res}$, which selects the ``pole strength'' $\sres$.  Equation~\ref{eq:Eres_to_B} can be used to map the tuning parameter to a laboratory $B$ field, given a value of $\delta \mu$ and $B_0$.  The $\sres$ parameter determined by selecting $A_{res}$ is also proportional to $\delta \mu$, since by definition (Eq. 35 of Ref.~\cite{Chin:2008})
\begin{equation}
\label{eq:sres}
 \sres = \frac {a_{bg} \Delta \delta \mu}{\bar{a}\bar{E}} \,.
\end{equation}
The 2-channel coupled Schr{\"o}dinger equation is solved numerically using standard methods and the bound state energies for $E<0$ and $s$-wave scattering length for $E \to 0$ are easily found as a function of $E_{res}$ and $\sres$ (or equivalently, $A_{res}$), given the fixed $a_{bg}$.

\begin{figure}
	 \includegraphics[width=1\columnwidth]{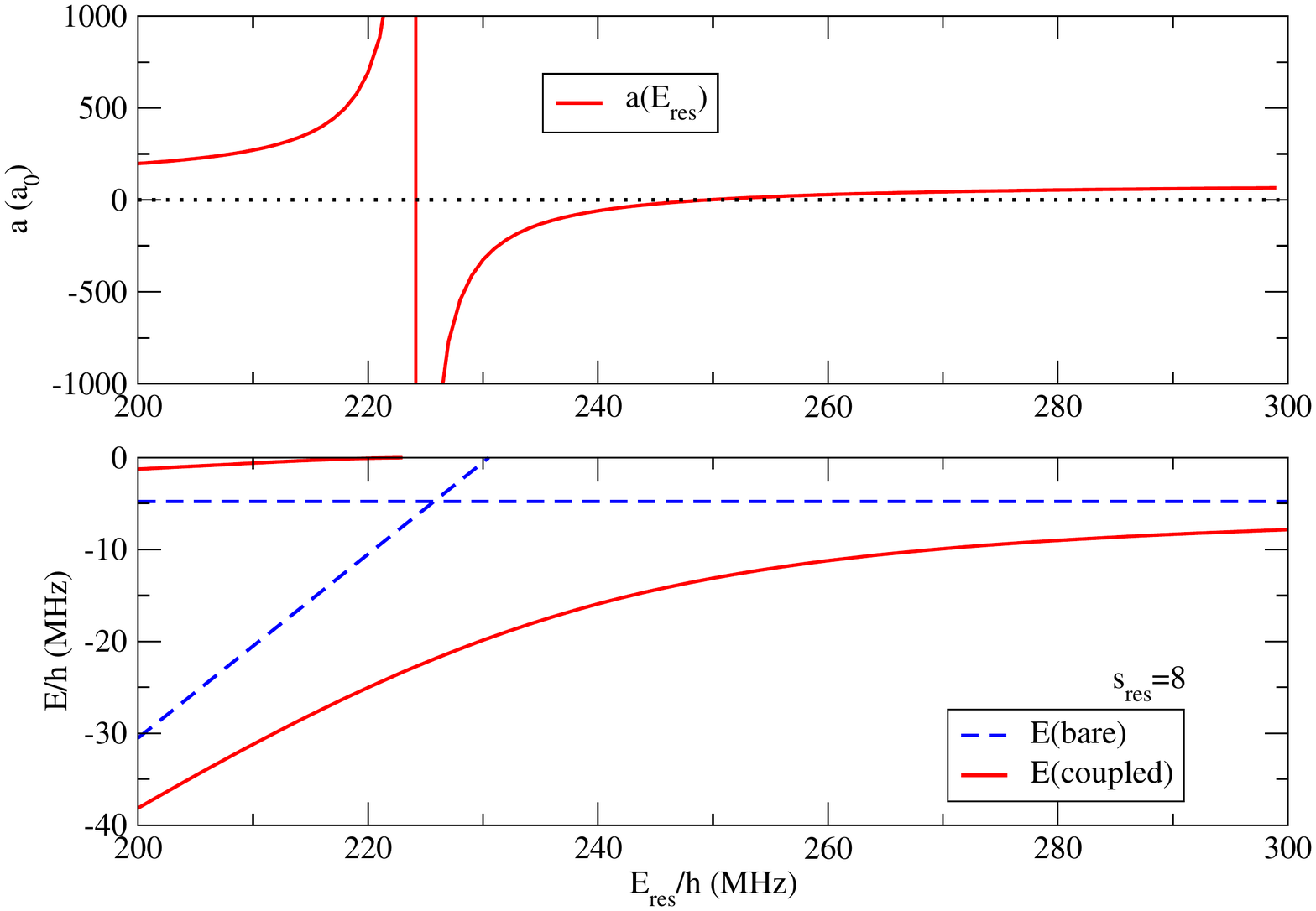}
	 \includegraphics[width=1\columnwidth]{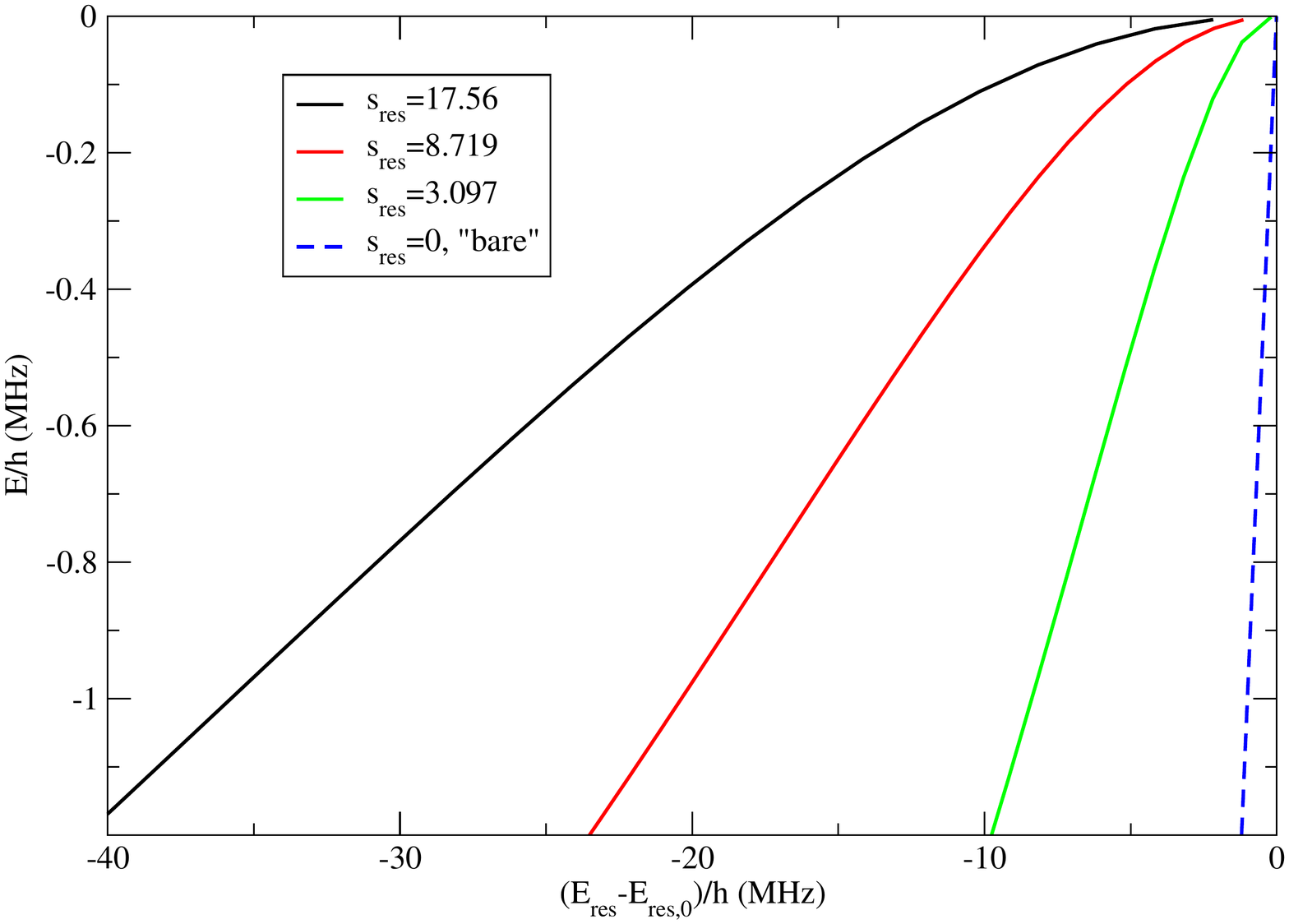}
	 \caption{(Upper panel pair) Scattering length and bound state energies versus $E_{res}$ for $\sres=$ 8.719.  The dashed lines show the bare (uncoupled) energies and the solid red lines show the result of the coupled calculation.  (Lower panel) Binding energy of the threshold bound state relative to its pole position at $E_{res}^0$ for three different values of pole strength $\sres$.  The dashed line shows the ramping bare level that generates the coupled halo bound state.
	  }
	 \label{fig:coupled_vs_Eres}
\end{figure}

Figure~\ref{fig:coupled_vs_Eres} shows examples of $E$ versus $E_{res}$ for different $\sres$ parameters.  The upper panel shows the bare states (uncoupled, $A_{res}=0$) and the coupled eigenstates for $\sres=$ 8.719.  The avoided crossing and the shift in pole from the bare crossing at $E_{res}^c/h=$230.51 MHz to the pole at $E_{res}^0/h=$ 224.14 MHz are evident.  The lower panel shows the binding energy of the threshold bound state versus $E_{res}-E_{res}^0$, that is, relative to its pole position, for several different values of $\sres$.  The width of the resonance increases linearly with $\sres$ (by definition from Eq.~\ref{eq:sres}).   In the $\sres \to 0$ limit, the binding energy curve would approach the linear curve for the bare state, with a small quadratic departure only very near threshold.  Keep in mind that the near-threshold $E$ versus $a$ curve becomes universal, that is, independent of $\sres$ when $\sres \gg 1$, and approaches the single channel curve, as indicated in Fig.~\ref{fig:E_vs_a}.  The universal $E(a)$ curve is independent of $\delta \mu$ when $\sres \gg 1$, in practice where $a/R_6 > 1$ and $\sres \gtrsim 10$.  Consequently, $a$ is determined by measuring $E$ in this universal limit (away from perturbations by narrow levels, of course).

\begin{figure}
	 \includegraphics[width=1\columnwidth]{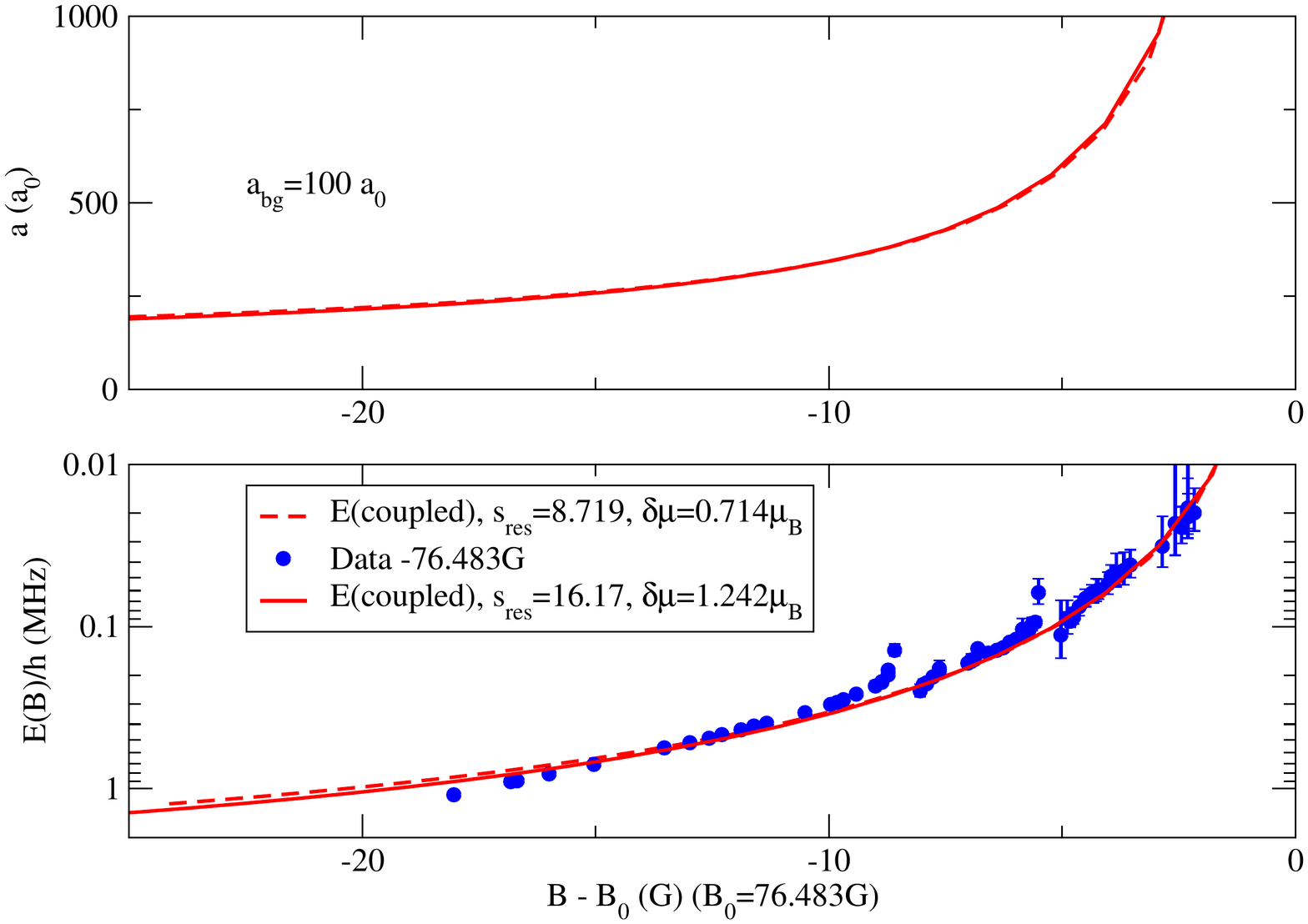}
	 \includegraphics[width=1\columnwidth]{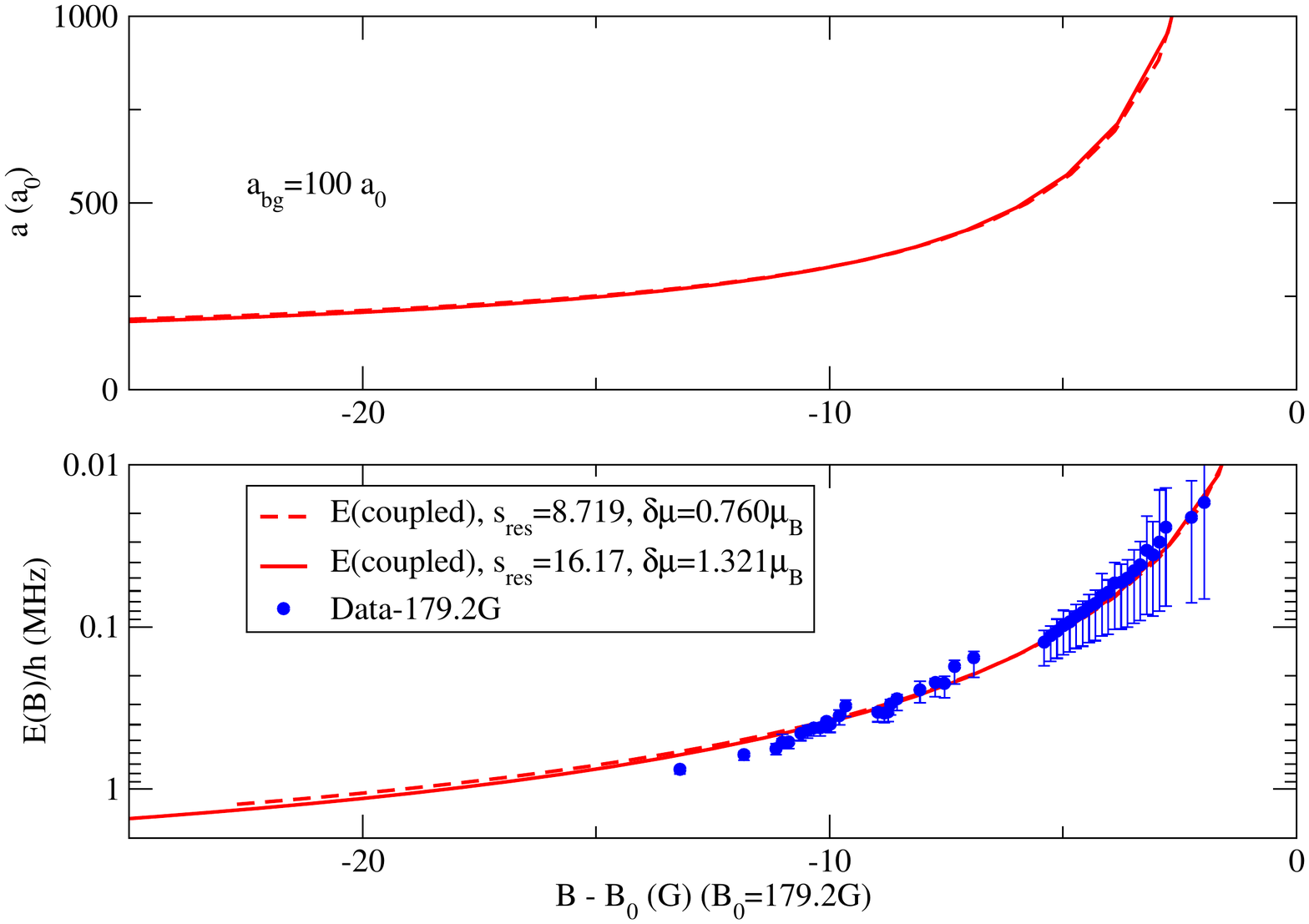}
	 \caption{(Upper Panel) Comparison of coupled channels model with the measured binding energies near the 76G resonance.  Very similar fits are obtained with different combinations of $\sres \gg 1$ and $\delta \mu$ such that $\sres/(\delta\mu)$ is constant.  (Lower Panel) Similar fit of the coupled channels model to the measured binding energies near the 170G resonance.
	  }
	 \label{fig:coupled_vs_B}
\end{figure}

Finally, Fig.~\ref{fig:coupled_vs_B} shows a comparison of a coupled channels calculation with the measured binding energies versus B for the two resonance regions.  The fit is not unique, since $\sres$ and $\delta \mu$ are not determined independently for such a halo state.  The figure shows fits with different assumed $\delta \mu$ and $\sres$ values.  Taking $\delta \mu =$ $g\mu_\mathrm{B}$ with $g=$ 1.24159 for Dy gives a reasonable minimum value to assume for the magnetic moment difference, namely that with the nearest atomic Zeeman level.  A good fit to the data is them obtained with $\sres=$ 16.17 for the 76G resonance.  Similar quality fits can be obtained for other combinations of $\sres$ and $\delta \mu$ such that $\sres/(\delta \mu)=(a_{bg}\Delta)/(\bar{a}\bar{E})$ is constant.  This determines $a_{bg}\Delta=$ 2600 a$_0$G and 2800 a$_0$G for the two fits for the 76G resonance.  The calculated width of the resonance with $\sres=$ 16.17 is $|E_{res}-E_{res}^0|/h=$ 47.11 MHz, or 27.1 G upon dividing by $\delta \mu = g\mu_\mathrm{B}$.  The fact that this value is consistent with the value of $\Delta =$ 31(6) G measured by examining the 3-body line shapes suggests that these values for $\sres$ and $\delta \mu$ are plausible.  A similar quality fit for the 170G resonance, shown in the figure, is obtained by taking the same $\sres$ values and assuming a $\delta \mu$ value 6 per cent larger.  Thus, the two resonances have comparable pole strengths, not surprising since they have quite similar measured $E$ versus $B$ curves.

\putbib[./FBHighField]
\end{bibunit}

\end{document}